\documentclass[12pt]{article}

\usepackage{amsmath}
\usepackage{mathrsfs}
\usepackage{xcolor}

\usepackage{draft} 
\usepackage{hyperref}
\usepackage{graphicx,color,subfig}
\usepackage{cite}
\usepackage{mciteplus}
\usepackage{skak}
\usepackage{empheq}
\usepackage{tikz}
\usepackage{booktabs}
\usepackage{siunitx}
\usepackage{bbm}
\usepackage{makecell}
\usepackage{float}

\usepackage{bm}
\usepackage{braket}
\usepackage{slashed}

\DeclareFontFamily{OT1}{pzc}{}
\DeclareFontShape{OT1}{pzc}{m}{it}{<-> s * [1.10] pzcmi7t}{}
\DeclareMathAlphabet{\mathpzc}{OT1}{pzc}{m}{it}

\newcommand{\wt}[1]{\widetilde{#1}}

\def\be#1\ee{\begin{align}#1\end{align}}

\makeatletter
\newcommand{\bdryno}{\mathpalette\bdry@no\relax}
\newcommand{\bdry@no}[2]{%
  \mspace{1mu}%
  \vbox{%
    \hbox{$\m@th#1\scriptstyle{\ast}$}
    \nointerlineskip
    \kern.25ex
    \hbox{$\m@th#1\scriptstyle{\ast}$}
    \kern-.06ex
  }%
  \mspace{1mu}%
}
\makeatother

\newcommand*\dif{\mathop{}\!\mathrm{d}}

\newenvironment{fix}{\color{red}}{\ignorespacesafterend}

\begin{document}

\unitlength = .8mm

\begin{titlepage}

\begin{center}

\hfill \\
\hfill \\
\vskip 1cm
\title{Supermoduli and PCOs at Genus Two}

\author{Charles Wang and Xi Yin}

\address{
Jefferson Physical Laboratory, Harvard University, \\
Cambridge, MA 02138 USA
}

\email{charles\_wang@g.harvard.edu, xiyin@fas.harvard.edu}

\abstract{We illustrate the relation between supermoduli integration and picture changing operators (PCOs) particularly concerning the role of vertical integration, in the context of superstring vacuum amplitudes, by an explicit comparison of different parameterizations of the supermoduli space of genus two super Riemann surfaces.
}

\vfill

\end{center}

\end{titlepage}

\eject

\begingroup
\hypersetup{linkcolor=black}

\tableofcontents

\endgroup

\section{Introduction}

Superstring perturbation theory, based on the path integral of worldsheet supergravity coupled to matter fields, has two incarnations: the super Riemann surface (SRS) formalism in which the scattering amplitude is expressed as an integral over the supermoduli space of SRS with punctures \cite{DHoker:1988pdl, Witten:2012ga}, and the picture changing operator (PCO) formalism in which the amplitude is expressed as an integral over the moduli space of ordinary Riemann surfaces with punctures and PCO insertions \cite{Friedan:1985ge, Verlinde:1987sd}. An important ingredient of the PCO formalism is the vertical integration prescription \cite{Sen:2014pia, Sen:2015hia}, which allows for evading spurious singularities. In a previous paper \cite{Wang:2022zad}, we explained how the vertical integration arises from a specific choice of the supermoduli integration contour, thereby establishing the equivalence between the two formalisms. In this follow-up paper, we will illustrate the general construction of \cite{Wang:2022zad} through the example of genus two vacuum amplitudes.\footnote{See \cite{DHoker:2001kkt, DHoker:2001qqx, DHoker:2001foj, DHoker:2001jaf, DHoker:2002hof, DHoker:2005dys, DHoker:2005vch, DHoker:2007csw, DHoker:2020prr, DHoker:2020tcq, DHoker:2021kks} for previous studies of genus two superstring amplitudes.}

Every genus two Riemann surface $\Sigma$ can be realized as a hyperelliptic curve, namely a double cover $\Sigma\to \mathbb{CP}^1$ branched at six points $x_1,\cdots, x_6$. Furthermore, such a branched covering of $\Sigma$ is unique up to the $\mathrm{PSL}(2, \mathbb{C})$ automorphism of $\mathbb{CP}^1$. This leads to a convenient parameterization of the bosonic moduli space ${\cal M}_2$, namely $\{x_1,\cdots,x_6\}\subset \mathbb{CP}^1$ modulo the action of $S_6$ permutation group and $\mathrm{PSL}(2, \mathbb{C})$. A spin structure $\epsilon$ on $\Sigma$ can be specified by partitioning the six branch points into two subsets $\{x_i^+\}$ and $\{x_j^-\}$. The spin structure is odd for a $(5,1)$ split and even for a $(3,3)$ split. The construction of a general super-Riemann structure over $(\Sigma,\epsilon)$ will be reviewed in section \ref{sec:hyper}. The corresponding supermoduli space will be denoted $\mathfrak{M}_{2,\epsilon}$.

In the PCO formalism, two holomorphic PCOs would need to be placed on $\Sigma$ in the absence of punctures. If we place both PCOs at the branch points of the majority type in the odd spin structure case (e,g. $x_1^+$ and $x_2^+$), or one PCO at a branch point of each type in the even spin structure case (e.g. $x_1^+$ and $x_1^-$), spurious singularities are avoided. However, due to monodromies that permute the branch points and constraints on the locations of PCOs near the boundary of the moduli space, there isn't a single assignment of the PCOs to branch points that is valid globally. Thus the interpolation between different fermionic directions in $\mathfrak{M}_{2,\epsilon}$ that arise from different PCO placements, and the corresponding vertical integration, must be considered.

We give an explicit parameterization of super coordinate patches on the moduli space of genus two SRS, that correspond to PCO insertions at branch points of the underlying hyperelliptic curve, for odd spin structure in section \ref{sec:odd} and even spin structure in section \ref{sec:evengenustwo}. We demonstrate the transition maps between super coordinates that correspond to different PCO arrangements, and their relations to vertical integration. In section \ref{sec:period}, we exhibit the relation between the super coordinates based on PCOs and those defined through the period matrix projection map in the genus two, even spin structure case.

We conclude with some remarks on applications and future perspectives in section \ref{sec:discuss}.

\section{SRS from hyperelliptic curves}
\label{sec:hyper}

\subsection{Hyperelliptic curves}

A genus $g$ hyperelliptic curve $\Sigma$ is defined as the locus
\ie\label{deffeq}
\big\{ (x, y)\in \mathbb{C}^2: y^2 = f(x)\big\}
\fe
where $f(x)$ is a degree $2g+2$ polynomial, completed by adding two points at infinity. A patch containing the points at infinity is parameterized by $(\wt{x},\wt{y})=(x^{-1}, x^{-g-1} y)$, subject to the equation
\begin{equation}\label{ftildeeq}
	\wt{y}^2 = \wt{f}(\wt{x}) \equiv \wt{x}^{2g+2} f\left(\wt{x}^{-1}\right).
\end{equation}
The curve admits an order two ``parity" automorphism that takes $(x, y)$ to $(x, -y)$. Therefore, functions on the curve can be split into even and odd parity components. 

A weight $k$ holomorphic form on $\Sigma$ can be described as $g(x) (\dif x)^k$ if parity even or $y g(x) (\dif x)^k$ if parity odd, where $g(x)$ is a meromorphic function with possibly poles at the branch points. 
Near infinity, we can write $\dif x = - \wt{x}^{-2} \dif \wt{x}$, and so 
\begin{equation}
	g(x) (\dif x)^k = g\left(\wt{x}^{-1}\right) (-1)^k \wt{x}^{-2k}(\dif \wt{x})^k.
\end{equation}
Therefore, $g(x)$ or $yg(x)$ depending on parity must decay at least as fast as $x^{-2k}$ as $x\to \infty$. Near the branch points, i.e. (simple) zeros of $f(x)$, we change the coordinate from $x$ to $y$, 
\begin{equation}
	g(x) (\dif x)^k = g(x) \left(\frac{2 y}{f'(x)}\right)^k (\dif y)^k.
\end{equation}
If $k=2j$ is even, then $g(x)$ can have an up to order $j$ pole at every branch point, regardless of parity. There are $2j(g-1)+1$ such linearly independent $g$'s in the even parity case, and $(2j-1)(g-1)+1$ in the odd parity case. If $k=2j+1$ is odd, in the even parity case $g(x)$ has up to order $j$ pole at every branch point, with a total number of $2j(g - 1) - 1$ modes; in the odd parity case $g(x)$ has up to order $j + 1$ pole at every branch point, and a total number of $(2j + 1)(g - 1) + 1$ modes. 


In particular, weight $k=2$ holomorphic forms, i.e. zero modes of $b$ ghost, are in 1-1 correspondence with cotangent vectors at the point $\Sigma$ in $\mathcal{M}_g$. There are $2g - 1$ parity even modes and $g - 2$ parity odd ones. 
In the genus two case, all three moduli deformations have even parity, preserving the hyperelliptic form of the curve. For higher genera, parity odd deformations destroy the hyperelliptic structure.


It will be useful to consider a moduli deformation of $\Sigma$ defined by cutting out a disc that contains a branch point $x_i$, and glue in a new disc parameterized by the coordinate $y_d$, with the transition map
\ie\label{yddef}
	y_d = y + \epsilon y^{-1} 
\fe
on the boundary of the disc, for infinitesimal $\epsilon$ with $\epsilon^2 = 0$.
Writing $f(x)\equiv (x-x_i) h(x)$, we could equivalently replace $y^{-1}$ on the RHS of (\ref{yddef}) by $y^{-1} {h(x)\over  f'(x_i)} =y^{-1}+{\cal O}(y)$, as the higher order terms in $y$ can be absorbed by a redefinition of the coordinate on the disc. Such a transition map can then be extended to the rest of the curve outside of the disc, giving the deformed equation 
\begin{equation}
	y_d^2 = y^2+2\epsilon {h(x)\over f'(x_i)} = \Big(x-x_i + {2\epsilon\over  f'(x_i)}\Big) h(x).
\end{equation}
Therefore, the moduli deformation (\ref{yddef}) is equivalent to moving the branch point from $x_i$ to $x_i - \frac{2 \epsilon}{f'(x_i)}$.

\subsection{Spin structure}

A spin structure on $\Sigma$ can be specified by choosing a factorization $f(x) = p(x) q(x)$, where $p(x)$ is a degree $n$ polynomial with $n - g - 1$ even, and $q(x)$ is a degree $m=2g+2-n$ polynomial. 
\ie\label{alphasplit}
	\alpha := \frac{y}{p(x)} = \frac{q(x)}{y}
\fe
is a parity odd meromorphic function on the curve, with a simple pole at every zero of $p(x)$, and simple zero at every zero of $q(x)$. 

We now promote $\Sigma$ to a complex supermanifold by introducing a fermionic coordinate $\eta$ where $p(x)\not=0$, and $\tau$ where $q(x)\not=0$, with the transition map
\ie\label{tauetarel}
\tau = \alpha \eta.
\fe
Near infinity, we pass to the coordinates $(\wt x,\wt y)$ that obey (\ref{ftildeeq}). Factoring $\wt{f} = \wt{p} \wt{q}$, where $\wt{p}(\wt{x}) = \wt{x}^n p(\wt{x}^{-1})$ and $\wt{q}(\wt{x}) = \wt{x}^m q(\wt{x}^{-1})$, we can then define the fermionic coordinates $\wt \eta$ and $\wt\tau$ related by
\begin{equation}
	\wt{\eta} = \wt{x}^{\frac{1}{2}(n - g + 1)}\eta,~~~ \wt{\tau} = \wt{x}^{\frac{1}{2}(m - g + 1)}\tau,
\end{equation}
with the transition map $\wt\tau=\wt\A\wt\eta$ between them, where $\wt{\alpha} = \wt{x}^{g + 1 - n} \alpha$. 
In particular, in the genus two odd spin structure case where $n = 1$ and $m = 5$, we have $\wt{\eta} = \eta$ and $\wt{\tau} = \wt{x}^{2} \tau$. In the even spin structure case, with
$n = m = 3$, we have $\wt{\eta} = \wt{x} \eta$ and $\wt{\tau} = \wt{x} \tau$.  Note that the spin structure is invariant under exchanging $p$ with $q$.

\begin{figure}[h!]
	\centering
	\scalebox{.8}{ 
		\begin{tikzpicture}
			\node[anchor=south west,inner sep=0] (image) at (0,0) {\includegraphics[width=.7\textwidth]{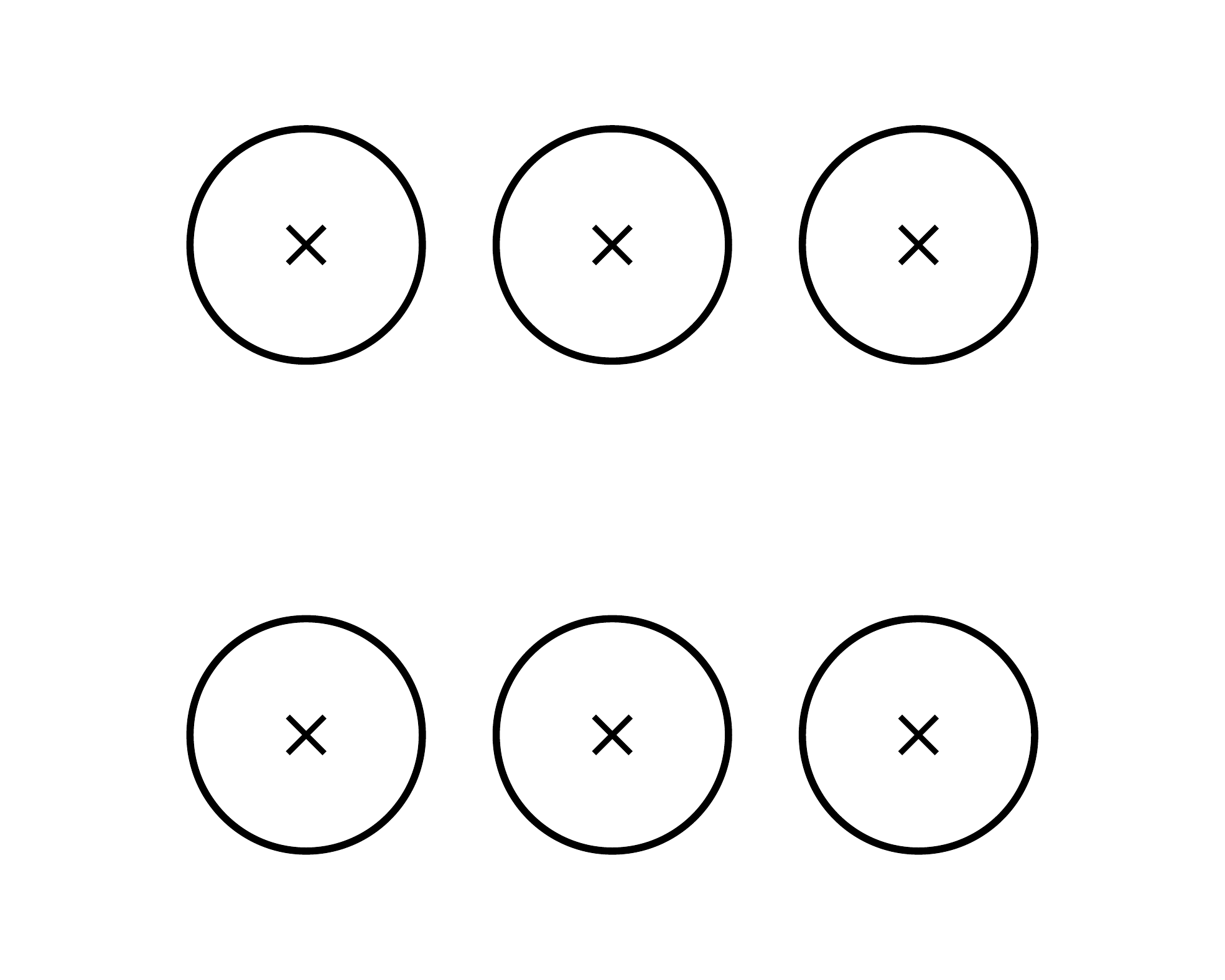}};
			\begin{scope}[shift={(image.south west)}, x={(image.south east)}, y={(image.north west)}]
				\node[anchor=north] at (0.25, 0.75 - 0.012) {\large $(y, \eta)$};
				\node[anchor=north] at (0.50, 0.75 - 0.012) {\large $(y, \eta)$};
				\node[anchor=north] at (0.75, 0.75 - 0.012) {\large $(y, \eta)$};
				\node[anchor=south] at (0.25, 0.25 + 0.012) {\large $(y, \tau)$};
				\node[anchor=south] at (0.50, 0.25 + 0.012) {\large $(y, \tau)$};
				\node[anchor=south] at (0.75, 0.25 + 0.012) {\large $(y, \tau)$};
				
				\node[anchor=south] (q) at (0.50, 0.9) {\large $q(x) = 0$};
				\draw [very thick, ->] (q) -- (0.25 + 0.03, 0.75 + 0.02);
				\draw [very thick, ->] (q) -- (0.75 - 0.03, 0.75 + 0.02);
				\draw [very thick, ->] (q) -- (0.50 + 0.00, 0.75 + 0.03);
				
				\node[anchor=north] (p) at (0.50, 0.1) {\large $p(x) = 0$};
				\draw [very thick, ->] (p) -- (0.25 + 0.03, 0.25 - 0.02);
				\draw [very thick, ->] (p) -- (0.75 - 0.03, 0.25 - 0.02);
				\draw [very thick, ->] (p) -- (0.50 + 0.00, 0.25 - 0.03);
				
				\node at (0.5, 0.5) {\large $(x, \eta)$};
			\end{scope}
 	\end{tikzpicture}}
 	\caption{
 		An illustration of coordinate charts on the super Riemann surface $\mathfrak{C}$. The circles represent boundaries of the discs containing the branch points of the underlying hyperelliptic curve, where $f(x) = 0$. Away from the branch points, the super coordinates are taken to be $x$ together with either $\eta$ or $\tau$. At the branch points, $x$ degenerates and $y$ is used instead. Near the branch point where $q(x) = 0$, the disc is parameterized by $(y, \eta)$ whereas $\tau$ degenerates. On the other hand, near the branch point where $p(x) = 0$, the roles of $\eta$ and $\tau$ are switched. 
 	}
\end{figure}
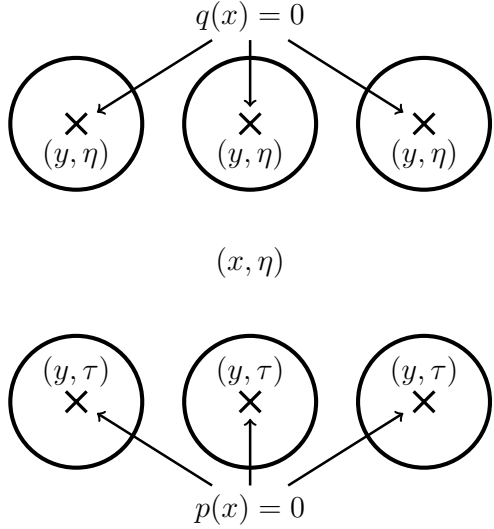

\subsection{Super-Riemann structure}

A super-Riemann structure can be specified by a totally non-integrable rank $0|1$ sub-bundle of the tangent bundle $T\Sigma$. In other words, there is a fermionic derivative operator $D$, defined locally up to rescaling such that $\{D, D\}$ is linearly independent from $D$ at every point. 
For instance, we can choose
\ie\label{ddsplit}
&	D_\eta = \partial_\eta + \alpha \eta \partial_x ,~~~~p(x)\not=0 ,
\\
&	D_\tau = \partial_\tau + \alpha^{-1}\tau \partial_x ,~~~~ q(x)\not=0.
\fe
Note that when both are defined, $D_\eta = \alpha D_\tau$. Also note that these are well defined at a branch point, say with $p(x)\not=0$, where $\alpha$ has a simple zero and $\partial_x$ has a simple pole, and we can write
\ie\label{detaexpr}
	D_\eta = \partial_\eta + \frac{f'(x)}{2 p(x)} \eta \partial_y.
\fe
Similarly, $D_\eta$ and $D_\tau$ are well-defined derivative operators at infinity. (\ref{ddsplit}) defines a split super-Riemann surface $\mathfrak{C}_0$ over $\Sigma$.

One should be cautious that (\ref{detaexpr}) is not of the standard form $D_\theta=\partial_\theta + \theta\partial_z$ associated with the canonical $(z, \theta)$ coordinates. Suppose we have a super coordinate chart $(z,\zeta)$ in which the super-Riemann structure corresponds to the fermionic derivative 
\ie\label{dzeta}
D_\zeta = \partial_\zeta + u(z) \zeta \partial_z,
\fe
then $D_\theta$ is related (up to overall rescaling) by $\theta = \sqrt{u(z)} \zeta$.
A superconformal transformation from $(z,\theta)$ to $(z',\theta')$ takes the standard form
\ie\label{sctrans}
& z' = f(z) + \theta g(z) h(z),
\\
& \theta' = g(z) + \theta h(z),~~~~ {\rm with}~ h^2 = \partial f + g\partial g.
\fe
In terms of the non-canonical coordinates $(z,\zeta)$ and $(z',\zeta')$, the latter associated with the fermionic derivative $D_{\zeta'}=\partial_{\zeta'} + \wt u(z')\zeta'\partial_{z'}$ where $\wt u(f(z))=u(z)$, the superconformal map (\ref{sctrans}) is expressed as
\ie\label{zetatrans}
& z' = f(z) + u(z)\, \zeta\, \wt g(z) h(z),
\\
& \zeta' = \wt g(z) + \zeta\, h(z),~~~~{\rm with} ~ h^2 = \partial f + u \wt g \partial \wt g,
\fe
where $\wt g(z) = {g(z)\over \sqrt{u(z)}}$. A useful formula for the Berezinian relating the integral form $[\dif z|\dif \zeta]$ to $[\dif z'|\dif \zeta']$ is
\ie\label{berf}
[\dif z'|\dif \zeta'] = {D_\zeta \left( u(z) \zeta' \right)\over u(z)}   [\dif z|\dif \zeta].
\fe
In later sections we will make frequent use of superconformal maps of the form (\ref{zetatrans}) to avoid proliferation of new coordinates.

\subsection{Automorphism and $\B$ zero modes}
\label{sec:auto}

The parity of $\Sigma$ lifts to an order 4 automorphism of the SRS $\mathfrak{C}_0$ that squares to $(-)^F$, where $F$ is the ``fermion number". This automorphism, which we will still refer to as parity, maps $(x,y,\eta)\mapsto (x,-y,i\eta)$ and $D_\eta\mapsto -i D_\eta$.


Let us examine the parity of weight ${3\over 2}$ holomorphic forms, which are in correspondence with zero modes of $\B$ ghost. Such a form of parity $-i$ can be written as the degree $1|1$ differential $y^{-1} g(x) \dif x \dif\eta$, where $g(x)$ is a holomorphic function away from the branch points. At a $q=0$ branch point, $y^{-1}$ has a simple pole, $\dif x\dif\eta$ has a simple zero, $g(x)$ is regular and the form is generically nonzero. At a $p=0$ branch point, $\dif x\dif \eta=\A^{-1} \dif x\dif\tau$ has a double zero, $g(x)$ must again be regular and the form has a simple zero. At infinity, $y$ grows as $x^{g+1}$, $\dif x\dif\eta$ grows as $x^{2+{1\over 2}(n-g+1)}$, and so $g(x)$ must be a polynomial of degree at most $g-1-{1\over 2}(n-g+1)$. 

In the genus two case, this means that for odd spin structure with $(n,m)=(1,5)$, we have two $\B$ zero modes of parity $-i$, both of them vanishing at the single $p=0$ branch point, and none of parity $+i$. For even spin structure with $(n,m)=(3,3)$, we have one $\B$ zero mode of each parity, $-i$ and $+i$, one of them vanishing at the $p=0$ branch points and the other vanishing at the $q=0$ branch points. 
If we are to place the two holomorphic PCOs at two of the branch points, say $x_1$ and $x_2$, to avoid spurious singularity we need the vectors $(\B(x_1),\B(x_2))$ for the two $\B$ zero modes to be linearly independent. This requires choosing $x_1, x_2$ to be among the 5 roots of $q(x)$ in the odd spin structure $(n,m)=(1,5)$ case, or  $q(x_1)=0$ and $p(x_2)=0$ in the even spin structure $(n,m)=(3,3)$ case.

\section{Genus two with odd spin structure}
\label{sec:odd}

We now give an explicit parameterization of the supermoduli space $\mathfrak{M}_{2,o}$ of a genus two SRS $\mathfrak{C}$ with odd spin structure, by connecting to the PCO formalism following the construction of \cite{Wang:2022zad}. The underlying ordinary Riemann surface $\Sigma$ is represented as a hyperelliptic curve (\ref{deffeq}), with $f(x)=p(x) q(x)$, where $p$ has degree 1 and $q$ has degree 5, and the spin structure is specified by the split of branch points according to (\ref{alphasplit}). 

\subsection{The PCO coordinates on $\mathfrak{M}_{2,o}$}

Starting with the split SRS $\mathfrak{C}_0$ described in the previous subsection, we can cut out a disc that contains the branch point $x=x_a$, and glue in the superdisc $D_a$ parameterized by $(w_a, \eta_a)$ with the transition map
\ie\label{srshypertransd}
& w_a = y - {f'(x)\over 2p(x)}\, \eta\, {f'(x) \nu_a\over y} ,
\\
& \eta_a = \eta - \frac{f'(x) \nu_a}{ y},
\fe
for $a=1,2$. Here $\nu_1, \nu_2$ are a pair of Grassmann-odd parameters. Note that (\ref{srshypertransd}) is of the required form (\ref{zetatrans}) for a superconformal transformation.\footnote{The fermionic derivative operator $D_{\eta_a} = \partial_{\eta_a} + {f'(u_a)\over 2p(u_a)}\eta_a \partial_{w_a}$, where $w_a^2=f(u_a)$, is related to to $D_\eta$ of (\ref{detaexpr}) up to rescaling.}
It is also equivalent to the standard disc gluing map given by (2.4) of \cite{Wang:2022zad} up to a constant rescaling of $\nu_a$ and a superconformal coordinate change on $D_a$. 

The gluing (\ref{srshypertransd}) defines a new SRS $\mathfrak{C}_\nu$. Furthermore, it gives an explicit parameterization of a super chart $\varphi: {\cal U}\times \mathbb{R}^{0|*2}\to \mathfrak{U}$ of the moduli space $\mathfrak{M}_{2,o}$, where the bosonic coordinates of ${\cal U}\subset {\cal M}_2$ can be taken as the branch points modulo ${\rm PSL}(2,\mathbb{C})$ action, and the fermionic coordinates are $(\nu_1, \nu_2)$. The projection $\pi: \mathfrak{U}\to {\cal U}$ simply forgets the fermionic coodinates. As explained in section 5.1 of \cite{Wang:2022zad}, integration over the fiber of $\pi$, i.e. in $\nu_1, \nu_2$, amounts to inserting a pair of PCOs at the branch points $x_1, x_2$.

\subsection{Transition map between PCO coordinates}
\label{sec:oddtransit}

Now consider a different family of SRS's $\mathfrak{C}'_{\nu}$, constructed as above but with $x_1$ replaced by another $q=0$ branch point $x_3$, and the corresponding super chart $\varphi': {\cal U}\times \mathbb{R}^{0|*2}\to \mathfrak{U}'$, along with the projection $\pi':\mathfrak{U}'\to {\cal U}$. Integration over the fiber of $\pi'$ amounts to inserting PCOs at $x_3$ and $x_2$. The transition map from $\mathfrak{U}$ to $\mathfrak{U'}$ takes the form 
\ie\label{transexplicit}
\varphi'^{-1}\circ \varphi : (x_i, \nu_1, \nu_2) \mapsto (x_i', \nu_1', \nu_2'),
\fe
where $x_i'$ are shifted branch point locations, that may in principle differ from $x_i$ by an (``infinitesimal") amount proportional to $\nu_1 \nu_2$. The fermionic coordinates of $\mathfrak{U}'$, denoted $\nu_1', \nu_2'$, are associated with the gluing maps analogous to (\ref{srshypertransd}) for the discs that contain $x_3'$ and $x_2'$ respectively. 

To find the explicit expression of the transition map (\ref{transexplicit}), let us begin by considering the following reparameterization of coordinates on $\mathfrak{C}_\nu$,
\ie\label{xyetamap}
& x' = x - \eta h(x) q(x) \nu_1,
\\
& y' = y - {f'(x)\over 2p(x)}\, \eta h(x) y \nu_1,
\\
& \eta' = \eta - h(x) y \nu_1,
\fe
with 
\ie\label{hexprs}
h(x) = {x_{12}x_{13}\over (x-x_1)(x-x_2)(x-x_3)}.
\fe
Passing from $\eta$ to $\tau$, one can see that (\ref{xyetamap}) is regular at the zero of $p(x)$. Regularity at infinity is ensured due to $h(x)$ decaying like $x^{-3}$. Thus (\ref{xyetamap}) is regular for $x\not\in \{x_1, x_2 ,x_3\}$. The expression for the fermionic derivative remains $D_{\eta'} = \partial_{\eta'} + \frac{y'}{p(x')} \partial_{x'}$.

\begin{figure}[h]
	\centering
	\scalebox{.8}{ 
		\begin{tikzpicture}
			\node[anchor=south west,inner sep=0] (image) at (0,5) {\includegraphics[width=\textwidth]{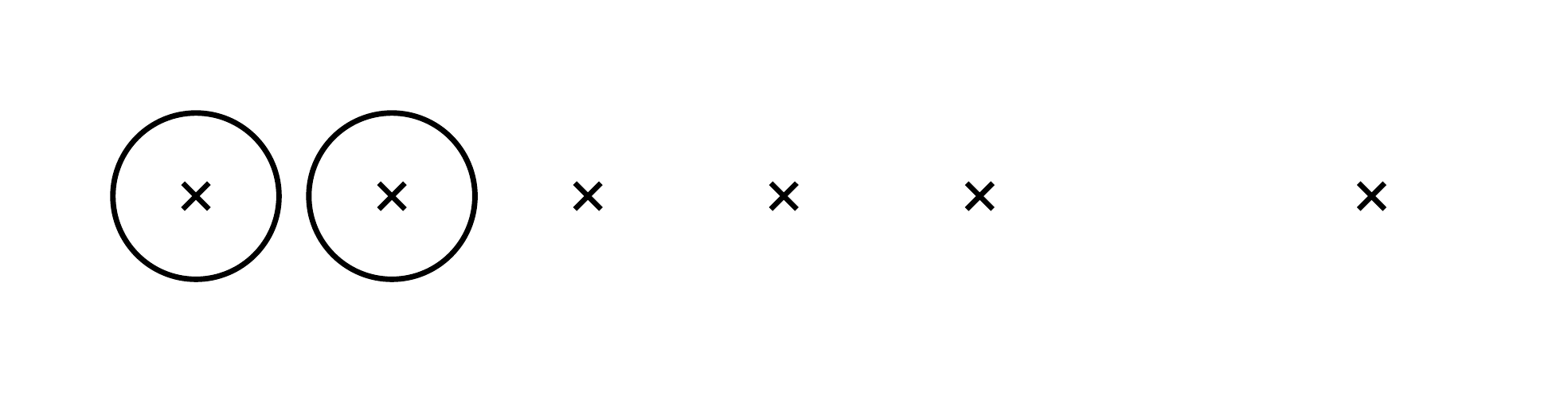}};
			\begin{scope}[shift={(image.south west)}, x={(image.south east)}, y={(image.north west)}]
				
				\node[anchor=south] (q) at (0.375, 0.8) {\large $q(x) = 0$};
				
				\node[anchor=south] (p) at (0.875, 0.8) {\large $p(x) = 0$};
				
				\node at (0.5, 0.3) {\large $(x, y, \eta)$};
				\node[anchor=north] at (0.125, 0.5) {$(w_1, \eta_1)$};
				\node[anchor=north] at (0.250, 0.5) {$(w_2, \eta_2)$};
				\node[anchor=south] at (0.125, 0.5 + 0.02) {$x_1$};
				\node[anchor=south] at (0.250, 0.5 + 0.02) {$x_2$};
				\node[anchor=south] at (0.375, 0.5 + 0.02) {$x_3$};
			\end{scope}
		\node[anchor=south west,inner sep=0] (image2) at (0,0) {\includegraphics[width=\textwidth]{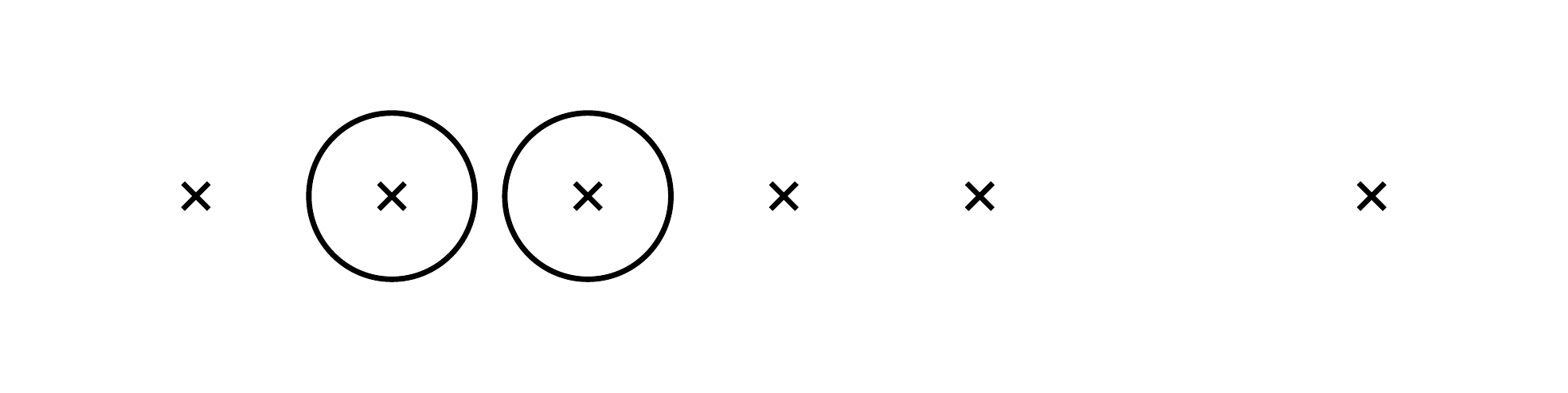}};
		\begin{scope}[shift={(image2.south west)}, x={(image2.south east)}, y={(image2.north west)}]
			
			\node[anchor=south] (q) at (0.375, 0.8) {\large $q(x) = 0$};
			\node[anchor=south] (p) at (0.875, 0.8) {\large $p(x) = 0$};
			
			\node at (0.5, 0.3) {\large $(x', y', \eta')$};
			\node[anchor=north] at (0.250, 0.5) {$(w_2, \eta_2)$};
			\node[anchor=north] at (0.375, 0.5) {$(y, \eta)$};
			\node[anchor=south] at (0.125, 0.5 + 0.02) {$x_1$};
			\node[anchor=south] at (0.250, 0.5 + 0.02) {$x_2$};
			\node[anchor=south] at (0.375, 0.5 + 0.02) {$x_3$};
		\end{scope}
	\end{tikzpicture}}
	\caption{The top and bottom diagrams illustrate two different sets of coordinates charts on the SRS $\mathfrak{C}_\nu$ in the odd spin structure case. The six branch points of the underlying hyperelliptic curve are marked with crosses, with the five $q(x) = 0$ branch points grouped together on the left, and the remaining $p(x) = 0$ branch point on the right. Top: the unprimed coordinate system becomes singular near the branch points $x_1, x_2$, whose neighborhoods are coordinatized by $(w_1, \eta_1)$ and $(w_2, \eta_2)$ respectively. Bottom: the primed coordinate system becomes singular near the branch points $x_2$ and $x_3$, whose neighborhoods are parameterized by $(w_2, \eta_2)$ and unprimed coordinates $(y, \eta)$. 
	}
\end{figure}
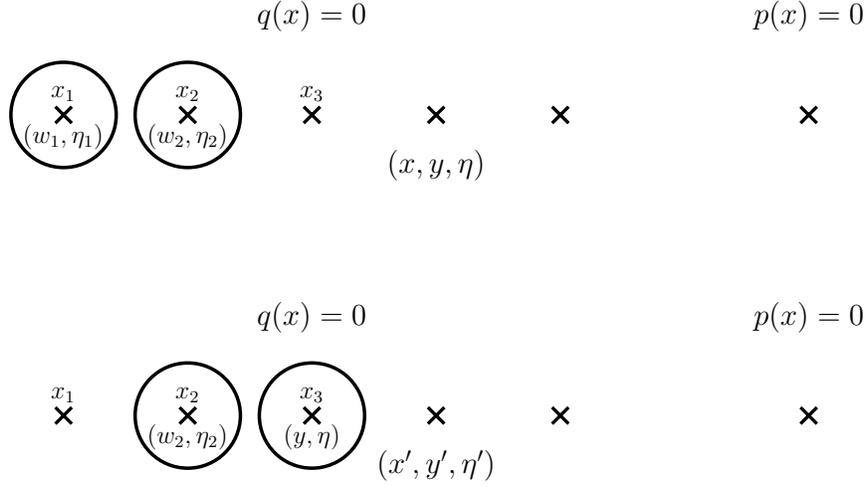

As $y'^2 = f(x')$ is still obeyed, the branch point positions are unchanged in the $x'$ coordinate. The gluing map (\ref{srshypertransd}) for the disc $D_1$, expressed in terms of the new coordinates $(y', \eta')$, is
\ie\label{extenddone}
& w_1 = y' + {f'(x')\over 2p(x')} \eta' \left[ h(x') - {f'(x')\over f(x')} \right] y' \nu_1,
\\
& \eta_1 = \eta' + \left[ h(x') - {f'(x')\over f(x')} \right] y' \nu_1 .
\fe
The RHS is regular at $x'=x_1$ by design of (\ref{hexprs}), and thus the coordinate system $(y', \eta')$ can be extended to $D_1$, erasing the said gluing map. 

On the other hand, we can view $(y, \eta)$ as the coordinates on a disc $D_1'$ containing $x_3$, whose relation to $(y', \eta')$ via (\ref{xyetamap}) amounts to the gluing map for $D_1'$. Comparing this to the analog of (\ref{srshypertransd}) that defines the coordinate map $\varphi'$ for the chart $\mathfrak{U}'$, we can read off
\ie
\label{eqn:odd_nu_1_prime}
\nu_1' = - h^{(-1)}(x_3)\nu_1 = {x_{12} \over x_{32}} \nu_1 .
\fe
Similarly, re-expressing the gluing map for $D_2$ in terms of $(y', \eta')$ gives, to first order in $\nu_1$ and $\nu_2$,
\ie{}
w_2 &= y' + {f'(x')\over 2p(x')} \eta' y' \left[ h(x') \nu_1 - {f'(x')\over f(x')} \nu_2\right] + {\cal O}(\nu_1\nu_2),
\\
\eta_2 &= \eta' + y' \left[ h(x')\nu_1 - {{f'(x')}\over {f(x')}} \nu_2  \right] + {\cal O}(\nu_1\nu_2).
\fe
From the residue at $x_2$ we read off 
\begin{equation}
	\label{eqn:odd_nu_2_prime}
	\nu_2' = \nu_2  -{\rm Res}_{x\to x_2} h(x)\nu_1 = \nu_2 + {x_{13}\over x_{23}}\nu_1.
\end{equation}
Therefore, the transition map (\ref{transexplicit}) takes the form
\ie
\label{eqn:odd spin transf prelim}
\left( x_i', \nu_1', \nu_2' \right) = \left(x_i + d_i \nu_1 \nu_2,  {x_{12} \over x_{32}} \nu_1,  \nu_2 + {x_{13}\over x_{23}}\nu_1 \right),
\fe
where the coefficients $d_i$ are yet to be determined functions of the $x_j$'s. 

Under the order 4 parity automorphism introduced in section \ref{sec:hyper}, the $x_i$'s are parity invariant, whereas $\nu_1, \nu_2$ are of the same imaginary parity $-i$. It follows that $\nu_1\nu_2$ is parity odd, while $d_i$ in (\ref{eqn:odd spin transf prelim}) is parity even, and therefore we must have $d_i=0$, leaving
\ie
\label{eqn:odd spin transf}
\left( x_i', \nu_1', \nu_2' \right) = \left(x_i,  {x_{12} \over x_{32}} \nu_1,  \nu_2 + {x_{13}\over x_{23}}\nu_1 \right).
\fe
The complete change of coordinates that identifies the SRS $\mathfrak{C}_\nu$ with $\mathfrak{C}'_{\nu'}$ is given in Appendix \ref{sec:odd spin explicit}. 

\subsection{The vanishing vertical integral}

As explained in \cite{Wang:2022zad}, the interpolation between the fermionic fibers of the projections $\pi:\mathfrak{U}\to {\cal U}$ and $\pi':\mathfrak{U}'\to {\cal U}$ considered in section \ref{sec:oddtransit} amounts to the vertical integration corresponding to moving a PCO from $x_1$ to $x_3$ on the underlying Riemann surface $\Sigma$. According to (\ref{eqn:odd spin transf}), however, the fermionic fibers of the two coordinate patches in fact agree (up to a linear change of fermionic coordinates $\nu_1, \nu_2$), and the interpolation is trivial. Indeed, we will see below that the vertical integration gives a vanishing result in this case.

The vertical integration in question produces a bosonic moduli integrand of the form
\ie\label{pcovert}
\left\langle e^{\cal B} \left[ \xi(x_3)-\xi(x_1) \right] \left[ {\cal X}(x_2) + d\xi(x_2)\right] \right\rangle_{\Sigma,\epsilon},
\fe
where we have omitted the anti-holomorphic operators insertions which do not affect the analysis here. The PCO ${\cal X}$ is given in the re-bosonized form by
\ie
{\cal X} = Q_B\cdot\xi = - {1\over 2} e^\phi G^{\rm matter} + c\partial\xi - {1\over 4} e^{2\phi} \partial \eta b - {1\over 4} \partial(e^{2\phi} \eta b).
\fe
Note that in the $(\phi,\xi,\eta)$ representation of $\B\C$-system correlators, there is implicitly an extra $\xi$-insertion of the form $\xi(x_*)$, where $x_*$ is an arbitrarily chosen point, that serves to absorb the unique zero mode of $\xi$ in its functional integration \cite{Friedan:1985ge}.

As we have chosen to parameterize the bosonic moduli deformations through those of the branch points $x_i$, ${\cal B}$ can be represented as
\ie
{\cal B} =\sum_i dx_i \oint_{C_{x_i}} {dy\over 2\pi i} {b^{(y)}(y)} \left(-{f'(x_i)\over 2y }\right) ,
\fe
where $C_{x_i}$ is a small circular contour enclosing $x_i$, and $b^{(y)}$ stands for the $b$ ghost defined in the conformal frame associated with the $y$-coordinate.

Due to the $bc$ and $\B\C$ ghost number anomaly, the only potentially non-vanishing part of (\ref{pcovert}) is
\ie\label{intsimp}
& \left\langle  e^{\cal B} \left[ \xi(x_3)-\xi(x_1) \right]  \left[ - {1\over 4} e^{2\phi} \partial \eta b(x_2) - {1\over 4} \partial(e^{2\phi} \eta b)(x_2) \right] \right\rangle_{\Sigma,\epsilon}
\\
&= \left\langle  e^{\cal B} \left[ \Theta(\B(x_1))-\Theta(\B(x_3)) \right]  \left[ {1\over 2} :\B\C \delta'(\B): b(x_2) - {1\over 4} \delta'(\B) \partial b(x_2) \right] \right\rangle_{\Sigma,\epsilon}.
\fe
This correlator is invariant under the parity automorphism $(x,y)\mapsto (x,-y)$ of $\Sigma$. On the other hand, under parity both zero modes of $\B(z)$ pick up a factor $-i$, and thus the functional integral for (\ref{intsimp}) picks up a minus sign. We conclude that (\ref{intsimp}) vanishes, as anticipated.

\section{Genus two with even spin structure}
\label{sec:evengenustwo}

\subsection{The PCO coordinates on $\mathfrak{M}_{2,e}$}

Next we turn to the supermoduli space $\mathfrak{M}_{2,e}$ of genus two SRS with even spin structure. The underlying hyperelliptic curve is given by (\ref{deffeq}) with $f(x)=p(x) q(x)$, where $p, q$ are both degree 3 polynomials, and the spin structure is specified as in (\ref{alphasplit}). As discussed at the end of section \ref{sec:hyper}, by placing one PCO at a branch point $x=x_1$ with $q(x_1)=0$, and another PCO at $x=x_2$ with $p(x_2)=0$, one constructs a SRS $\mathfrak{C}_\nu$ with the gluing map
\ie\label{wetapmasev}
& w_1 = y - {f'(x)\over 2p(x)} \eta {f'(x)\nu_1\over y},
\\
& \eta_1 = \eta - {f'(x)\nu_1\over y},
\fe
for a super disc $D_1$ parameterized by $(w_1,\eta_1)$ that contains the branch point at $x_1$, and
\ie\label{wetapmasevqta}
& w_2 = y - {f'(x)\over 2q(x)} \tau {f'(x)\nu_2\over y},
\\
& \tau_2 = \tau - {f'(x)\nu_2\over y},
\fe
for $D_2$ paramererized by $(w_2, \tau_2)$ that contains $x_2$. Note that in writing (\ref{wetapmasevqta}) we have passed to the fermionic coordinate $\tau= y^{-1} q(x) \eta$ which is non-singular at $x=x_2$. The family of SRS's parameterized by the branch points and $\nu_1, \nu_2$ defines a coordinate patch $\varphi: {\cal U}\times \mathbb{R}^{0|*2}\to \mathfrak{U}$ of the supermoduli space.

\subsection{Transition map between PCO coordinates}
\label{sec:eventaspc}

Similarly, we can construct another family of SRS $\mathfrak{C}'_\nu$, and the corresponding supermoduli patch $\varphi': {\cal U}\times \mathbb{R}^{0|*2}\to \mathfrak{U}'$, by replacing $D_1$ with a disc $D_1'$ that contains another branch point $x_3$ where $q(x_3)=0$, whose gluing map still takes the form (\ref{wetapmasev}).

As $\eta$ has parity $i$, $\tau$ has parity $-i$, $\nu_1$ and $\nu_2$ have parity $-i$ and $i$ respectively. It follows that transition map between $\mathfrak{U}$ and $\mathfrak{U}'$, of the form (\ref{transexplicit}), must
be such that
\ie\label{xandnuas}
x_i' = x_i + d_i \nu_1 \nu_2,~~~ \nu_1' = a_1 \nu_1, ~~~\nu_2' = a_2 \nu_2,
\fe
where $d_i$ and $a_1, a_2$ are functions of the $x_j$'s. 
%
%

To proceed, consider the following superconformal reparameterization of the coordinates on $\mathfrak{C}_\nu$,
\ie\label{xyrtansev}
&x' = x + \left(\frac{1}{x - x_3} - \frac{1}{x - x_1}\right)  q(x) \eta \nu_1 ,
~~~~~y' = y + \left(\frac{1}{x - x_3} - \frac{1}{x - x_1}\right)  {f'(x)\over 2p(x)} y \eta \nu_1 ,
\\ 
&\eta' = \eta + \left(\frac{1}{x - x_3} - \frac{1}{x - x_1}\right) y \nu_1 ,
~~~~~ \tau' = \tau + \left(\frac{1}{x - x_3} - \frac{1}{x - x_1}\right)q(x)\nu_1.
\fe
This transformation is regular for $x\not\in \{x_1, x_3\}$, including at infinity as well as at the other branch points, and is constructed so that the transition map between the coordinates $(w_1, \eta_1)$ on the disc $D_1$ (\ref{wetapmasev}) and $(y',\eta')$ can be extended regularly to $x=x_1$, thereby erasing the gluing map for $D_1$. On the other hand, it introduces a nontrivial transition map for a disc $D_1'$ containing $x_3$, parameterized by $(y,\eta)$, from which we can read off $\nu_1'=\nu_1$, or $a_1=1$ in (\ref{xandnuas}).

Near $x_2$, the transition map that expresses the coordinates $(w_2, \tau_2)$ on the disc $D_2$ (\ref{wetapmasevqta}) in terms of $(y',\tau')$ is
\ie{}
& w_2 = y' - {(f'(x'))^2 \tau'\nu_2\over 2q(x') y'} + {1\over 2}f'(x') {x_{31}\over (x'-x_3)(x'-x_1)} \left[ { f'(x')\over y'} \nu_1\nu_2 - \tau'\nu_1 \right],
\\
&\tau_2 = \tau' - {f'(x') \nu_2 \over y'} + {x_{31}\over (x'-x_3)(x'-x_1)} \left[ \left( f''(x') - {(f'(x'))^2\over 2 f(x')} \right) \tau' \nu_1\nu_2 - q(x') \nu_1 \right].
\fe
We can put this back in the standard form of the gluing map (\ref{wetapmasevqta}) by a further change of coordinates from $(y',\tau')$ to $(y'',\tau'')$, with
\ie{}
& y'' = y' + {(f'(x'))^2\over y'} {x_{31}\over (x'-x_3)(x'-x_1)} \nu_1\nu_2, \\
& \tau'' = \tau' - {\left(f'(x')\right)^2 \over 2 f(x')} {x_{31}\over (x'-x_3)(x'-x_1)} \tau' \nu_1 \nu_2, 
\fe
as well as a change of coordinates on $D_2$ from $(w_2,\tau_2)$ to $(w_2'',\tau_2'')$ defined by
\ie{}
& w''_2 = w_2 + {f'(x') \over 2}  {x_{31}\over (x'-x_3)(x'-x_1)} \tau' \nu_1, \\
& \tau''_2 = \tau_2 + t'(x')\left[q(x') \nu_1 - f''(x')\tau_2 \nu_1 \nu_2\right].
\fe
\begin{figure}[h]
	\centering
	\scalebox{.8}{ 
		\begin{tikzpicture}
			\node[anchor=south west,inner sep=0] (image) at (0,6) {\includegraphics[width=\textwidth]{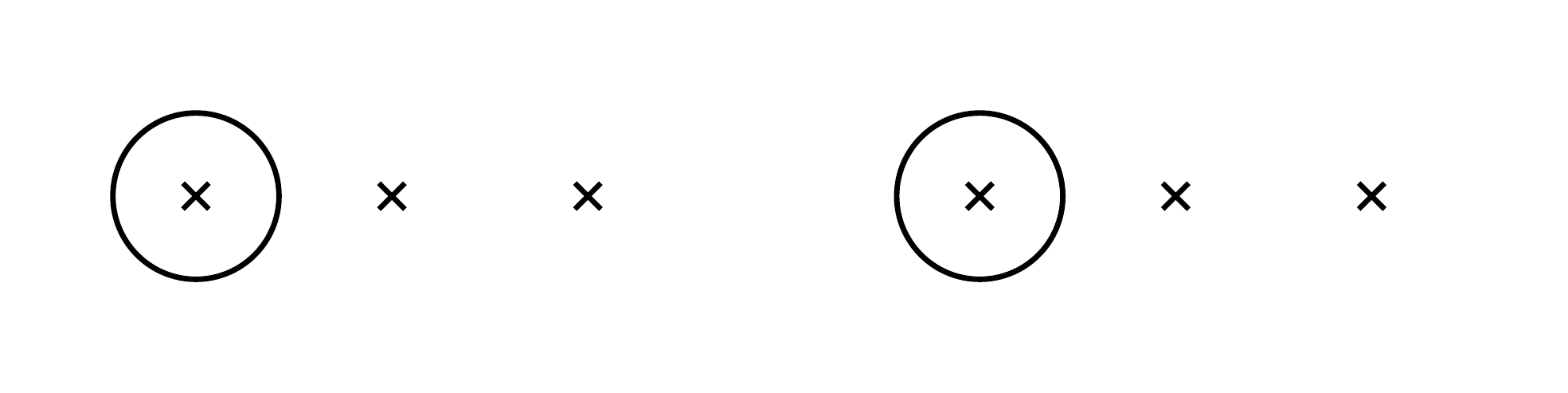}};
			\begin{scope}[shift={(image.south west)}, x={(image.south east)}, y={(image.north west)}]
				
				\node[anchor=south] (q) at (0.250, 0.8) {\large $q(x) = 0$};
				
				\node[anchor=south] (p) at (0.750, 0.8) {\large $p(x) = 0$};
				
				\node at (0.5 - 0.015, 0.5) {\large $(x, y, \eta)$};
				\node[anchor=north] at (0.125, 0.5) {$(w_1, \eta_1)$};
				\node[anchor=north] at (0.625, 0.5) {$(w_2, \eta_2)$};
				\node[anchor=south] at (0.125, 0.5 + 0.02) {$x_1$};
				\node[anchor=south] at (0.625, 0.5 + 0.02) {$x_2$};
				\node[anchor=south] at (0.250, 0.5 + 0.02) {$x_3$};
			\end{scope}
			\node[anchor=south west,inner sep=0] (image2) at (0,3) {\includegraphics[width=\textwidth]{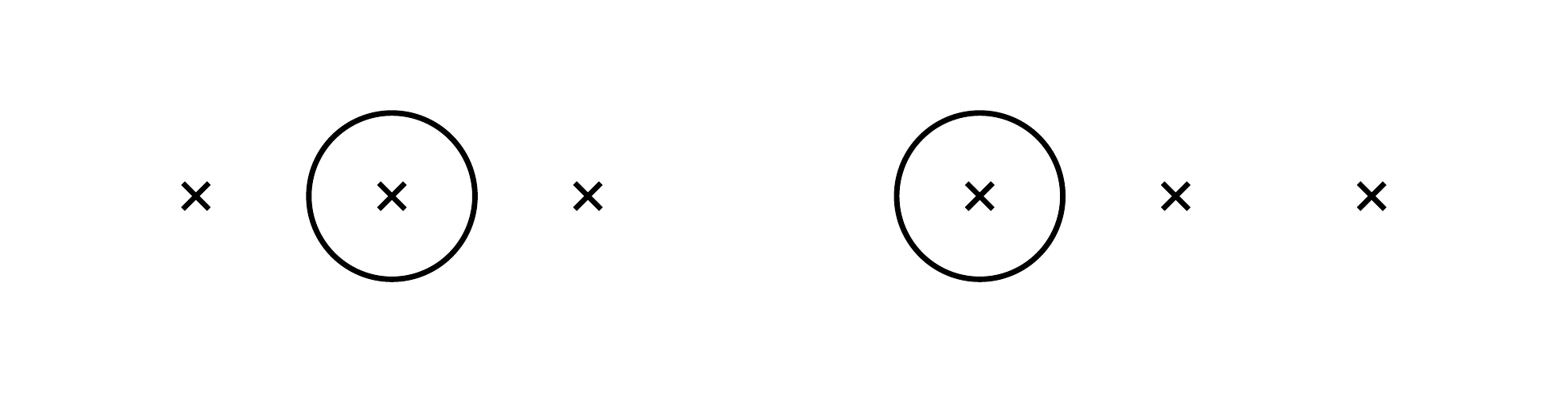}};
			\begin{scope}[shift={(image2.south west)}, x={(image2.south east)}, y={(image2.north west)}]
				
				\node at (0.5 - 0.015, 0.5) {\large $(x', y', \eta')$};
				\node[anchor=north] at (0.625, 0.5) {$(w_2, \eta_2)$};
				\node[anchor=north] at (0.250, 0.5) {$(y, \eta)$};
				\node[anchor=south] at (0.125, 0.5 + 0.02) {$x_1$};
				\node[anchor=south] at (0.625, 0.5 + 0.02) {$x_2$};
				\node[anchor=south] at (0.250, 0.5 + 0.02) {$x_3$};
			\end{scope}
		
			\node[anchor=south west,inner sep=0] (image3) at (0,-0.5) {\includegraphics[width=\textwidth]{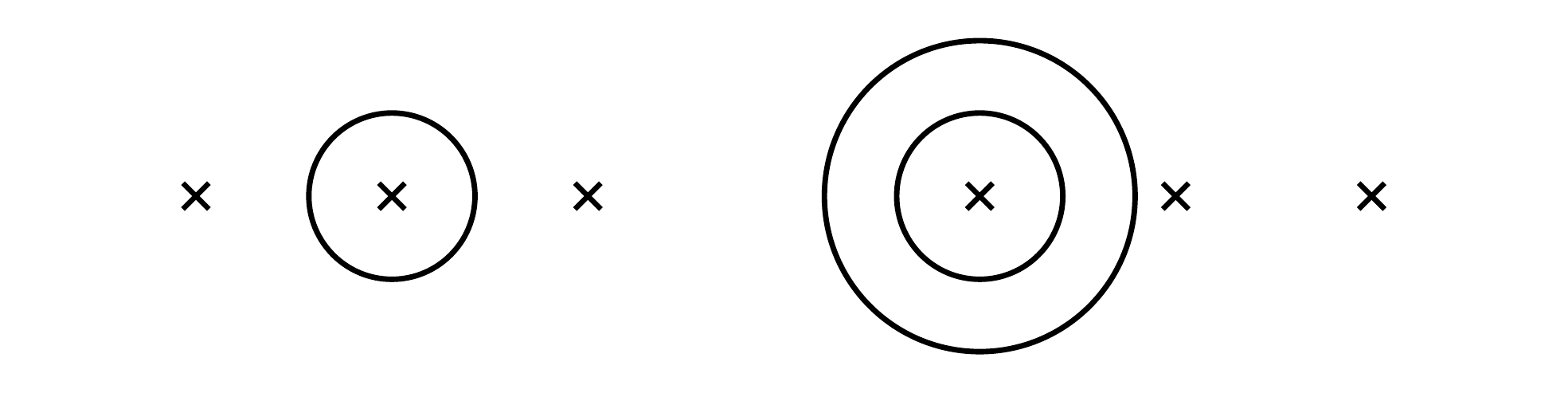}};
			\begin{scope}[shift={(image3.south west)}, x={(image3.south east)}, y={(image3.north west)}]
				
				\node at (0.45, 0.3) {\large $(x', y', \eta')$};
				\node[anchor=north] at (0.625, 0.5) {$(w_2'', \eta_2'')$};
				\node[anchor=north] at (0.625, 0.3) {$(y'', \eta'')$};
				\node[anchor=north] at (0.250, 0.5) {$(y, \eta)$};
				\node[anchor=south] at (0.125, 0.5 + 0.02) {$x_1$};
				\node[anchor=south] at (0.625, 0.5 + 0.02) {$x_2$};
				\node[anchor=south] at (0.250, 0.5 + 0.02) {$x_3$};
			\end{scope}
	\end{tikzpicture}}
	\caption{ The top, middle, and bottom diagrams illustrate the unprimed, primed, and double-primed coordinate systems on $\mathfrak{C}_\nu$ in the even spin structure case. The six branch points are marked with crosses, with the three $q(x) = 0$ branch points grouped together on the left, and the other three $p(x) = 0$ branch points on the right. Top: the unprimed coordinate system, with $(w_1, \eta_1)$ and $(w_2, \eta_2)$ parameterizing the discs containing $x_1$ and $x_2$ respectively. Middle: the primed coordinate system, which is regular away from $x_2$ and $x_3$. Bottom: The double-primed coordinate system on an annulus surrounding $x_2$, whose transition maps into the disc coordinates $(w_2'',\eta_2'')$ take the standard form. 
	}
\end{figure}
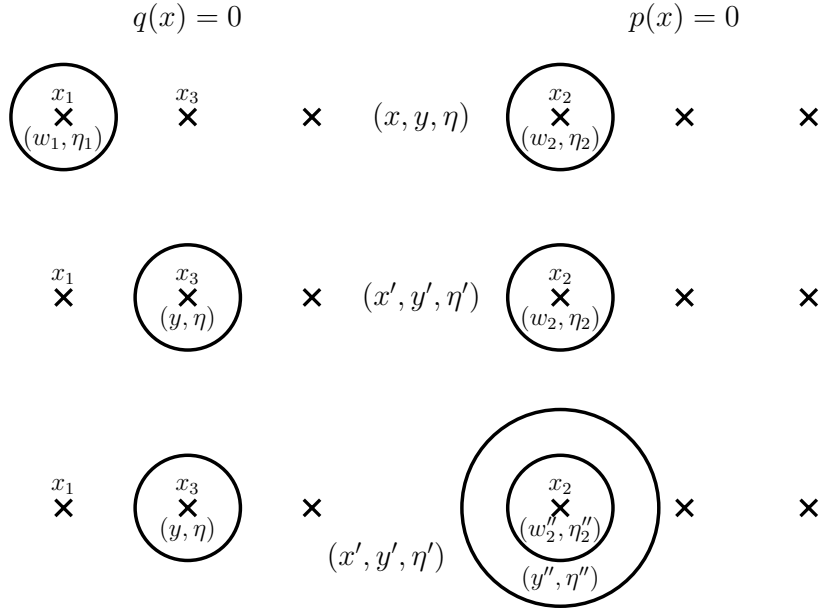
We conclude that $\nu_2'=\nu_2$, or $a_2=1$ in (\ref{xandnuas}).
Up to regular terms at $(x',y') = (x_2,0)$, the transformation between $y''$ and $y'$ is of the form (\ref{yddef}), and leads to a shift of of the branch point $x_2$ to
\ie
x_2' = x_2 - 2 f'(x_2) {x_{31}\over x_{23}x_{21}} \nu_1\nu_2.
\fe 
In conclusion, the transition map from $\mathfrak{U}$ to $\mathfrak{U}'$ takes $(x_2, \nu_1, \nu_2)$ to
\ie\label{phiphitrans}
(x_2', \nu_1',\nu_2') = \left( x_2 - 2 f'(x_2) {x_{31}\over x_{23}x_{21}} \nu_1\nu_2, \nu_1, \nu_2 \right),
\fe
while leaving all other branch points $x_i$ ($i\not=2$) invariant.

A subtlety arises when moving both PCOs, say from $x_1, x_2$ to $x_3, x_4$, where $q(x_3)=p(x_4)=0$. Following the above procedure, if we first move one PCO from $x_1$ to $x_3$, and then the other PCO from $x_2$ to $x_4$, the final branch point positions are
\ie\label{branchfinala}
& x_1' = x_1 ,~~~~ x_2' = x_2 - 2 f'(x_2) {x_{31}\over x_{23}x_{21}} \nu_1\nu_2, ~~~~ x_3' = x_3 + 2 f'(x_3){x_{42}\over x_{34}x_{32}} \nu_1\nu_2,~~~ x_4'=x_4.
\fe
If we exchange the order of the two moves, the final branch point positions are
\ie\label{branchfinalb}
& \wt x_1' = x_1 + 2 f'(x_1){x_{42}\over x_{14}x_{12}} \nu_1\nu_2,~~~~ \wt x_2'=x_2,~~~~\wt x_3'=x_3 ,~~~~ \wt x_4' = x_4 - 2 f'(x_4) {x_{31}\over x_{43}x_{41}} \nu_1\nu_2.
\fe
However, the resulting SRS should be independent of the order of the PCO moves. Indeed, the branch points $x_i'$ appearing in (\ref{branchfinala}) and $\wt x_i'$ in (\ref{branchfinala}) are related by an infinitesimal ${\rm PSL}(2,\mathbb{C})$ transformation
\ie
\widetilde x = x -  {2 x_{31} x_{42} f(x) \over (x-x_1)(x-x_2)(x-x_3)(x-x_4)} \nu_1\nu_2,
\fe
which leaves the remaining two branch points $x_5, x_6$ invariant, as expected.

\subsection{The corresponding vertical integral}
\label{sec:verteven}

The vertical integral corresponding to moving the PCO from $x_1$ to $x_3$ produces an integrand of the form (\ref{pcovert}), or simply (\ref{intsimp}), except that the spin structure $\epsilon$ is even as specified in section \ref{sec:eventaspc}. The correlators in question can be obtained from a more general correlator of the $(\phi,\xi,\eta)$ system of the form
\ie\label{xietasam}
\left\langle \xi(z_1) \xi(z_2) e^{\phi}(u) e^\phi \eta(v) \right\rangle.
\fe
Here we work in the large Hilbert space and it is necessary to keep track of the dependence on the position of the extra $\xi$ insertion \cite{Friedan:1985ge}. 

Using the general formula for the correlators of $\xi,\eta$ and $e^{\A\phi}$ in terms of theta functions and prime forms \cite{Verlinde:1987sd}, one deduces that the spurious poles of (\ref{xietasam}) are the same as those of $\left\langle \delta(\B(z_1)) \delta(\B(u)) \right\rangle$ and $\left\langle \delta(\B(z_2)) \delta(\B(u)) \right\rangle$. There are further OPE poles as $v\to z_1, z_2, u$, an OPE zero as $z_1\to z_2$, and extra ``spurious zeros" that occur at the same loci as the spurious poles of $\left\langle \delta(\B(z_1)) \delta(\B(z_2)) \delta(\B(u)) \delta(\C(v)) \right\rangle$. In the limit $u\to v$, in particular, the spurious poles of (\ref{xietasam}) are the same as those of
\ie
{\left \langle \delta(\B(z_1)) \delta(\B(u)) \right\rangle \left\langle \delta(\B(z_2)) \delta(\B(u)) \right\rangle \over \left\langle \delta(\B(z_1)) \delta(\B(z_2)) \right \rangle}
\fe
Specializing to the hyperelliptic curve $\Sigma$, we will be interested in the limit $u,v\to x_2$, where $p(x_2)=0$. We will further take $z_2$ to be another branch point $x_4$ with $p(x_4)=0$. As seen in section \ref{sec:auto}, one of the $\B$ zero modes vanishes at the $p=0$ branch points, the other vanishes at the $q=0$ branch points. It follows that both $\left \langle \delta(\B(x)) \delta(\B(x_2)) \right\rangle$ and $\left \langle \delta(\B(x)) \delta(\B(x_4)) \right\rangle$ have simple poles at precisely the roots of $p(x)$. Consequently, the following correlators\footnote{The superscript ``$(y)$" indicates that the operator is defined in the conformal frame associated with the $y$-coordinate, which is regular at the branch point.}
\ie\label{fgepscorr}
& F(x) = \left\langle \xi(x) (e^{2\phi} \eta)^{(y)}(x_2) \xi(x_4) \right\rangle^{\B\C}_{\Sigma,\epsilon},
\\
& G(x) = \left\langle \xi(x) (\partial\phi e^{2\phi} \eta)^{(y)}(x_2) \xi(x_4) \right\rangle^{\B\C}_{\Sigma,\epsilon},
\\
& H(x) = \left\langle  \xi(x) (e^{2\phi} \partial\eta)^{(y)}(x_2) \xi(x_4)  \right\rangle^{\B\C}_{\Sigma,\epsilon}
\fe
have no spurious poles in $x$; their only poles in $x$ are due to the OPE singularity as $x\to x_2$. Combining with the single-valuedness in $x$, and regularity at infinity,\footnote{For instance, the OPE singularity indicates that $F(x)$ must be proportional to ${y\over x-x_2}$, which cannot be regular at infinity unless its coefficient vanishes.} we conclude that
\ie{}
& F(x)=G(x)=0,
\\
& H(x) = {1\over f'(x_2)(x-x_2)} \left\langle (e^{2\phi})^{(y)} (x_2) \xi(x_4)  \right\rangle^{\B\C}_{\Sigma,\epsilon} + H(\infty).
\fe
We can compare this with
\ie{}
K(x) &= \left \langle \delta(\B)^{(y)}(x) \delta(\B)^{(y)}(x_2) \right\rangle = {(f'(x))^{3\over 2}\over \sqrt{p(x) q(x_2) (f'(x_2))^3 }} \left\langle (e^{2\phi})^{(y)} (x_2) \xi(x_4)  \right\rangle^{\B\C}_{\Sigma,\epsilon},
\fe
and deduce
\ie{}
H(x_3)-H(x_1)& = {1\over 2} (f'(x_2))^2 {x_{13}\over x_{32}x_{12}} \left[ {2 p(x_1)\over (f'(x_1))^3} \right]^{1\over 2}  \left[ {2 q(x_2)  \over  (f'(x_2))^3} \right]^{1\over 2} K(x_1)
\\
&= {1\over 2} ( f'(x_2))^2{x_{13}\over x_{32}x_{12}}  \left[ {2p(x_3)\over (f'(x_3))^3} \right]^{1\over 2}  \left[ {2q(x_2)  \over  (f'(x_2))^3} \right]^{1\over 2} K(x_3).
\fe
The result of vertical integration thus contributes to the moduli integrand
\ie\label{vertevenres}
\left\langle e^{\cal B} \left[ - 2(f'(x_2))^2 {x_{13}\over x_{32}x_{12}} b(x_2) \right] \left[ {2 p(x_1)\over (f'(x_1))^3} \right]^{1\over 2}\delta(2\B(x_1))^{(y)} \left[ {2 q(x_2)  \over  (f'(x_2))^3} \right]^{1\over 2} \delta(2\B(x_3))^{(y)} \right\rangle_{\Sigma,\epsilon}.
\fe
This is precisely the result of integrating out the fermionic parameters $\nu_1, \nu_2$ over the interpolation between the fibers of $\pi: \mathfrak{U}\to {\cal U}$ and $\pi': \mathfrak{U}'\to {\cal U}$ related by the bosonic shift  (\ref{phiphitrans}), as a special case of the consideration of section 5 of \cite{Wang:2022zad}.

\section{Comparison with period matrix projection}
\label{sec:period}

The moduli space $\mathfrak{M}_{2,e}$ of genus two SRS with even spin structure admits a projection onto its reduced space via the period matrix \cite{DHoker:1988pdl, DHoker:2002hof}. The latter gives rise to a natural coordinate system on $\mathfrak{M}_{2,e}$ in terms of the bosonic moduli and fermionic coordinates parameterizing the fiber of the projection. In this subsection we explain the relation between the period matrix projection and the PCO coordinates.

Recall that the period matrix of an ordinary Riemann surface is defined through the integrals of a basis of holomorphic 1-forms along 1-cycles. On a SRS $\mathfrak{C}$, one can consider a weight ${1\over 2}$ integral form $J$, written in a canonical local coordinate system $(z,\theta)$ as
\ie\label{jsrs}
J = (k(z) + \theta j(z)) [\dif z | \dif \theta].
\fe
On a split SRS $\mathfrak{C}_0$, $j(z)$ is a weight 1 form and $k(z)$ a weight ${1\over 2}$ form. In the case of even spin structure, there are no global weight ${1\over 2}$ forms, and (\ref{jsrs}) reduces to the weight 1 form $j$ on the underlying ordinary Riemann surface. The period matrix defined by integrals of (\ref{jsrs}) along cycles can be extended to a general non-split SRS. In the case of odd spin structure, on the other hand, there is a mode for $k(z)$ which interferes with the extension of this notion of period matrix to non-split SRSs.

We henceforth focus on the case of a genus two SRS with even spin structure. Consider $\mathfrak{C}_\nu$ constructed by deforming from the split SRS with the gluing maps (\ref{wetapmasev}), (\ref{wetapmasevqta}), corresponding to PCOs placed at the branch points $x_1, x_2$. In $(x, \eta)$ coordinates, the integral form $J$ can be written as
\ie\label{supercurrent}
J &= (B + \eta C) [dx|d\eta] .
\fe
It is shown in Appendix \ref{appen:period} that $B, C$ are constrained to be of the form
\ie{}
& B = - {a(x_1) \nu_1\over x-x_1} - {a(x_2) \nu_2\over x-x_2}  \A^{-1},
\\
& C = y^{-1}  a(x) + y^{-1} \left[ - {f'(x_1) a(x_2)\over 2 x_{12} (x-x_1)} - {f'(x_2) a(x_1)\over 2 x_{12}(x-x_2)} + \wt c(x) \right] \nu_1\nu_2,
\fe
where $a(x)$ and $\wt c(x)$ are linear polynomials in $x$.

The fiber of the period matrix projection over $\Sigma$, specified by the branch point locations $x_i$, consists of SRSs of the form
\ie\label{fiberphi}
\varphi (x_i + s_i(x_j) \nu_1\nu_2, \nu_1,\nu_2),
\fe
where $\varphi:{\cal U}\times \mathbb{R}^{0|*2}\to \mathfrak{U}$ is the coordinate map of the supermoduli space defined in section \ref{sec:evengenustwo}. The deformation of the branch point locations $\delta x_i=s_i \nu_1\nu_2$ is such that we can find a corresponding deformed weight ${1\over 2}$ form $J+\delta J$, whose periods are independent of $\nu_1,\nu_2$. The calculation of Appendix \ref{appen:period} gives
\ie\label{sresulta}
&s_1 = \frac{f'(x_1)}{x_{12}} - A(x_1) ,
\\
&s_2 = \frac{f'(x_2)}{x_{12}} - A(x_2),
\\
& s_i = -A(x_i),~~~i=3,4,5,6,
\fe
where $A(x)$ is a degree 3 polynomial whose leading coefficient is fixed to be $A^{(3)} = {f^{(6)}} x_{12}$. Similarly to the discussion in section \ref{sec:eventaspc}, the shift due to the quadratic part of $A(x)$ can be undone by an infinitesimal ${\rm PSL}(2,\mathbb{C})$ transformation.

Alternatively, we can represent the fiber (\ref{fiberphi}) of the period matrix projection in terms of the coordinate map $\varphi'$, defined with one PCO moved from $x_1$ to $x_3$ as in section \ref{sec:evengenustwo}, as
\ie\label{fiberphipr}
\varphi' (x_i + \wt s_i(x_j) \nu_1\nu_2, \nu_1,\nu_2),
\fe
where
\ie{}
&\wt s_3 = \frac{f'(x_3)}{x_{32}} - \wt A(x_3) ,
\\
&\wt s_2 = \frac{f'(x_2)}{x_{32}} - \wt A(x_2),
\\
& \wt s_i = -\wt A(x_i),~~~i=1,4,5,6,
\fe
and $\wt A(x)$ is a degree 3 polynomial with $\wt A^{(3)} = {f^{(6)}} x_{32}$. Comparing (\ref{fiberphi}) with (\ref{fiberphipr}), we see that the transition map $\varphi'^{-1}\circ\varphi$ takes $(x_i,\nu_1,\nu_2)$ to $(x_i+\delta x_i,\nu_1,\nu_2)$, with
\ie\label{delxis}
\delta x_i = (\wt s_i(x_j) - s_i(x_j))\nu_1\nu_2.
\fe
In particular, if we choose the quadratic terms of $\wt A(x)$ to be such that 
\ie
\wt A(x) = A(x) + {f(x) x_{31}\over (x-x_1)(x-x_2)(x-x_3)}, 
\fe
the only non-vanishing shift in (\ref{delxis}) is $\delta x_2 = - 2 f'(x_2) {x_{31}\over x_{12} x_{32}}$, in agreement with (\ref{phiphitrans}).

\section{Discussions}
\label{sec:discuss}

Let us summarize the results of this paper. We explicitly parameterized the supermoduli space $\mathfrak{M}_{2,\epsilon}$ of a genus two SRS $\mathfrak{C}$, with either odd or even spin structure $\epsilon$, by identifying the fermionic moduli $\nu_1, \nu_2$ with deformation parameters of gluing maps for a pair of discs cut out around a pair of branch points $x_1, x_2$ of the underlying hyperelliptic curve $\Sigma$. Spurious singularities are avoided provided that $x_1, x_2$ are branch points of the appropriate types specified by the spin structure. The resulting super coordinates on $\mathfrak{M}_\epsilon$, which we refer to as ``PCO coordinates", is closely related to the insertion of a pair of PCOs at the branch points in the PCO formalism. Said equivalently, the PCO coordinates specifies a projection $\pi$ from a supermoduli chart $\mathfrak{U}$ of $\mathfrak{M}_{2,\epsilon}$ to its reduced space ${\cal U}$, $\pi: \mathfrak{U}\to {\cal U}$, whose fiber is parameterized by the fermionic coordinates $(\nu_1, \nu_2)$. 

Crucially, the super chart $\mathfrak{U}$ and the projection $\pi$ depend on the choice of branch points $x_1, x_2$, We shall emphasize this dependence with the notations $\mathfrak{U}_{x_1,x_2}$ and $\pi_{x_1, x_2}$. Changing $x_1$, for instance, to another branch point $x_3$ of the same (admissible) type, results in a different chart $\mathfrak{U}_{x_3,x_2}$ along with its projection $\pi_{x_3, x_2}$. We have explicitly computed the transition map between the two sets of super coordinates on the overlap $\mathfrak{U}_{x_1,x_2}\cap \mathfrak{U}_{x_3,x_2}$, given by (\ref{eqn:odd spin transf}) in the case of odd spin structure, and (\ref{phiphitrans}) in the case of even spin structure.

The results for the transition maps indicate that the (fermionic) fibers of $\pi_{x_1, x_2}$ and $\pi_{x_3, x_2}$ in fact coincide in the odd spin structure case, but differ in the even spin structure case. In the context of genus two superstring vacuum amplitudes, the supermoduli integration can be performed by integrating out the fermionic fibers of the projections defined over a set of cells of the reduced space, while making up for the mismatch between the fibers of adjacent cells by a set of interpolation integrals. The latter was shown in \cite{Wang:2022zad} to reproduce the vertical integration prescription \cite{Sen:2014pia, Sen:2015hia} in the PCO formalism. Here we see explicitly that, as the PCO configuration is changed from $\{x_1, x_2\}$ to $\{x_3, x_2\}$, the corresponding vertical integral vanishes in the odd spin structure case, and is nontrivial in the even spin structure case. The latter produces a correction to the bosonic moduli integrand (\ref{vertevenres}) that is precisely equivalent to the interpolation integral between the fibers of $\pi_{x_1,x_2}$ and $\pi_{x_3, x_2}$.

In the even spin structure case, there is a global projection of the supermoduli space defined by the period matrix, which results in a super coordinate system that is different from the PCO coordinates. The explicit transition map between the period matrix coordinates and the PCO coordinates is determined in (\ref{fiberphi}) with (\ref{sresulta}).

While the exercise performed in this paper primarily serves to illustrate the general construction of \cite{Wang:2022zad}, it reveals a number of interesting features of the supermoduli space $\mathfrak{M}_2$ that may be useful for explicit computations of superstring amplitudes, with arbitrary worldsheet matter SCFT and any types of GSO projection. For instance, the explicit construction of non-singular PCO coordinates based on the branch points and the corresponding vertical integrals is applicable for the evaluation of genus two free energy of the 2D type 0 string theory at finite temperature, that would serve as a highly nontrivial test of the conjectured duality with the matrix quantum mechanics \cite{Takayanagi:2003sm, Douglas:2003up, Balthazar:2022atu, Balthazar:2022apu}. It is also straightforward to generalize the analysis in this paper to include NS punctures, which may provide a convenient setup for computing $n$-point genus two superstring amplitudes for both odd and even spin structures (see \cite{DHoker:2020prr, DHoker:2020tcq, DHoker:2021kks} for recent progress on the subject).

\section*{Acknowledgements}

This work is supported in part by a Simons Investigator Award from the Simons Foundation, by the Simons Collaboration Grant on the Non-Perturbative Bootstrap, and by DOE grant DE-SC0007870.

\appendix

\section{Details of coordinate transformations on the SRS with odd spin structure}
\label{sec:odd spin explicit}


In this appendix we give further details of the superconformal coordinate transformation that identifies the SRS $\mathfrak{C}_\nu$ and $\mathfrak{C}_{\nu'}'$ of section \ref{sec:oddtransit}. 

The original coordinates on $\mathfrak{C}_\nu$ consist of (two out of) the four variables $x, y, \eta, \tau$ subject to the conditions (\ref{deffeq}) and (\ref{tauetarel}).
The fermionic derivatives $D_\eta$ and $D_\tau$ that define the super-Riemann structure satisfy 
\ie
\label{eqn: srs hyperelliptic differentials}
	&D_\eta \eta = 1, ~~~~ D_\eta x = \alpha \eta, ~~~~~~~ D_\eta y = \frac{f'(x)}{2 p(x)} \eta,\\
	&D_\tau \tau = 1, ~~~~ D_\tau x = \alpha^{-1} \tau, ~~~~ D_\tau y = \frac{f'(x)}{2 q(x)} \tau, \\
	&D_\tau = \alpha^{-1} D_\eta.
\fe
These coordinates are regular everywhere except near the branch points $x_1$ and $x_2$, where the disc coordinates 
\begin{equation}
	\label{eqn: disk coordinates}
	\left(w_a, \eta_a\right) = \left(y - \frac{f'(x)}{2 p(x)}\nu_a, \eta - \frac{f'(x)}{y} \nu_a\right)
\end{equation}
respectively for $a=1,2$, are regular instead. 
The fermionic derivatives $D_{\eta_a}$ on the discs satisfy
\ie
	&D_{\eta_a}\eta_a = 1, ~~~~ D_{\eta_a}w_a = \frac{f'(x)}{2 p(x)} \eta_a,\\
	&D_{\eta_a} = \left(1 + \frac{y}{p(x)}\partial_x\left(\frac{f'(x)}{y}\right)\eta\nu_a\right) D_\eta.
\fe
Inverting (\ref{eqn: disk coordinates}), we can write 
\ie
	&y = w_a + \frac{\left(f'(x_a)\right)^2}{2 p(x_a)}\frac{1}{w_a} \eta_a \nu_a + {\cal O}(w_a \eta_a \nu_a) \\
	&\eta = \eta_a + \frac{f'(x_a)}{w_a} \nu_a + {\cal O}(w_a \nu_a).
\fe
Note that $x$ is regular at $w_a = 0$, 
\begin{equation}
	x = x_a + \frac{f'(x_a)}{2 p(x_a)} \eta \nu_a + {\cal O}(w_a^2).
\end{equation}
Next, we introduce the primed coordinates as in (\ref{xyetamap}). They satisfy relations identical to those of in  (\ref{deffeq}), (\ref{tauetarel}), (\ref{eqn: srs hyperelliptic differentials}), with the unprimed variables replaced by the primed variables. For instance, the fermionic derivative $D_\eta$ is replaced by $D_{\eta'}$, related by
\begin{equation}
	D_{\eta'} = \left(1 + \partial_x \left(h(x) y\right) \frac{y}{p(x)} \eta \nu_1\right) D_\eta,
\end{equation}
where $h(x)$ is given in (\ref{hexprs}).
The primed coordinates are regular everywhere except at the branch points $x_2$ and $x_3$. Note that regularity at $x_1$ comes from cancelation of poles. Near $x_3$, the singular part of the transformation between $(y,\eta)$ viewed as coordinates on the disc $D_1'$ containing $x_3$, and $(y', \eta')$ as coordinates outside of $D_1'$, is 
\ie
	y' &= y + \frac{\left(f'(x_3)\right)^2}{2 p(x_3)} \frac{1}{y} \eta \nu_1' + {\cal O}(y \eta \nu_1), \\
	\eta' &= \eta + \frac{f'(x_3)}{y} \eta'_1 + {\cal O}(y \eta_1),
\fe
where $\nu_1'$ is defined in (\ref{eqn:odd_nu_1_prime}). This is the same form as that associated with a PCO insertion. 
At $x_2$, the singular part of the relevant coordinate transformation is
\ie{}
	&y' = w_2 + \frac{\left(f'(x_2)\right)^2}{2 p(x_2)} \frac{1}{w_2} \eta_2 \nu_2' + \frac{\left(f'(x_2)\right)^3}{2 p(x_2)}h^{(-1)}(x_2) \frac{1}{w_2^2} \nu_1 \nu_2 + {\cal O}(\nu_1 \nu_2 , w_2 \eta_2 \nu_1, w_2 \eta_2 \nu_2),
	 \\
	&\eta' = \eta_2 + \frac{f'(x_2)}{w_2} \nu_2' + \frac{w_2}{2 p(u_2)} f'(u_2)  h'(u_2) \eta_2 \nu_1 \nu_2 + {\cal O}(w_2\nu_1 , w_2 \nu_2),
\fe
where $\nu'_2$ is as in (\ref{eqn:odd_nu_2_prime}), and $u_2$ is the analog of the $x$-coordinate on the disc, related to $w_2$ by $w_2^2=f(u_2)$. This matches (\ref{eqn: disk coordinates}) up to regular terms and terms proportional to $\nu_1 \nu_2$. 
To put this in the standard form of the disc gluing map associated with a PCO insertion, we need to perform a further bosonic coordinate redefinition  
\ie{}
	&x'' = x' + r(x') y' \nu_1 \nu_2,
	\\
	&y'' = y' + \frac{1}{2} f'(x') r(x') \nu_1 \nu_2,
\fe
where
\begin{equation}
	r(x) = - \frac{f'(x_2)}{p(x_2)} h^{(-1)}(x_2)\frac{1}{x - x_2} = q'(x_2) \frac{x_{13}}{x_{23}} \frac{1}{x - x_2}.
\end{equation}
Note that $D_{\eta'}$ acts on $x''$ as
\begin{equation}
	D_{\eta'}x'' = \frac{y''}{p(x'')}\left(1 + \frac{y''}{p(x'')}\partial_{x''}\left(r(x'') p(x'')\right)\nu_1 \nu_2\right) \eta'.
\end{equation}
Now introducing the double-primed fermionic coordinates and the fermionic derivative
\ie{}
	&\eta'' = \left(1 + \frac{y''}{2 p(x'')}\partial_{x''}\left(r(x'') p(x'')\right)\nu_1 \nu_2\right)\eta', \\
	&\tau'' = \left(1 + \frac{y''}{2 q(x'')}\partial_{x''}\left(r(x'') q(x'')\right)\nu_1 \nu_2\right)\tau', \\
	&D_{\eta''} = \left(1 - \frac{y''}{2 p(x'')}\partial_{x''}\left(r(x'') p(x'')\right)\nu_1 \nu_2\right)D_{\eta'},
\fe
we find that $(x'',y'',\eta'',\tau'')$ satisfy relations identical to (\ref{deffeq}), (\ref{tauetarel}), and (\ref{eqn: srs hyperelliptic differentials}), where all variables are replaced with their double-primed counterparts. 

The double-primed coordinates are still regular away from $x_2$ and $x_3$. The singularity of these coordinates at $x_3$ remains the same as that of the single-primed coordinates. At $x_2$, on the other hand, the singularity of the double-primed coordinates is given by 
\ie\label{yetadoubpr}
	&y'' = w_2 + \frac{f'(x_2)}{2 p(x_2)}\frac{1}{w_2} \eta_2 \nu_2' + {\cal O}(\nu_1 \nu_2 , w_2 \eta_2 \nu_1, w_2 \eta_2 \nu_2), \\
	&\eta'' = \eta_2 + \frac{f'(x_2)}{w_2} \nu_2' + \frac{w_2}{2 p(u_2)}\left(f'(u_2) h'(u_2)  + \partial_{u_2} \left(r(u_2) p(u_2)\right)\right) \eta_2 \nu_1 \nu_2 + {\cal O}(w_2\nu_1 , w_2 \nu_2). 
\fe
where $\nu'_2$ is given by (\ref{eqn:odd_nu_2_prime}), and $u_2$ obeys $w_2^2=f(u_2)$. 
One can verify that (\ref{yetadoubpr}) can be put in the standard form (\ref{eqn: disk coordinates}) by a regular redefinition of the coordinates $(w_2,\eta_2)$ on the disc $D_2$.\footnote{Note in particular that the term proportional to $h'(u_2)$ in the second line of (\ref{yetadoubpr}), which is singular at the center of the disc, can be absorbed by a redefinition of $w_2$ that is regular on the disc.}
In conclusion, the map
\ie{}
	&x'' = x - \frac{q(x)}{2} \eta h(x) \nu_1 + r(x) y \nu_1 \nu_2, \\
	&y'' = y - \frac{f'(x)}{2 p(x)} \eta h(x) y \nu_1 + \frac{1}{2} f'(x) r(x) \nu_1 \nu_2,\\
	&\eta'' = \eta - h(x) y \nu_1 + \frac{1}{2} \frac{y}{p(x)} \partial_x \left(r(x) p(x)\right)\eta \nu_1 \nu_2, \\
	&\tau'' = \frac{y''}{p(x'')} \eta'' 
\fe
together with its extension to the discs gives the desired isomorphism between $\mathfrak{C}_\nu$ and $\mathfrak{C}_{\nu'}'$.

\section{Calculating the fiber of the period matrix projection}
\label{appen:period}

We would like to identify a SRS $\mathfrak{C}_\nu$ defined via (\ref{wetapmasev}) and (\ref{wetapmasevqta}), that has the same period matrix as a split SRS $\mathfrak{C}'$. The integral form $J$ (\ref{supercurrent}) on $\mathfrak{C}_\nu$ is constrained by parity to be of the form
\ie\label{jexpr}
J = \left(\nu_1 b_1(x) + \nu_2 \alpha^{-1} b_2(x) + \eta y^{-1} \left(a(x) + c(x) \nu_1 \nu_2\right) \right)[\dif x|\dif \eta].
\fe
Regularity demands that $a(x)$ is linear in $x$. At infinity, $b_1(x)$ and $b_2(x)$ must decay as $x^{-1}$, whereas $c$ must grow at most linearly in $x$.
Near the branch point $x_1$, we pass to the coordinates $(w_1, \eta_1)$ (\ref{wetapmasev}), with (using the Berezinian formula (\ref{berf}))
\ie\label{wetachfir}
	[\dif w_1| \dif \eta_1] = \left(1 -  \eta \partial_y\left( {(f'(x))^2\over 2p(x) y}\right) \nu_1\right) [\dif y| \dif \eta],
\fe
and
\ie{}
	\eta_1 [\dif w_1| \dif \eta_1] = \left(\eta - \frac{f'(x) \nu_1}{y}\right) [\dif y|\dif \eta] = \frac{f'(x)}{2}\left(- \frac{f'(x)}{f(x)} \nu_1 + \frac{\eta}{y}\right) [\dif x| \dif \eta].
\fe
The regularity of (\ref{jexpr}) at $x_1$ thus requires $b_1(x)$ to have a pole at $x_1$ with residue $b_1^{(-1)}(x_1) = - a(x_1)$. Near $x_2$, a similar analysis with $\eta$ and $\tau$ flipped gives $b_2^{(-1)}(x_2) = - a(x_2)$. Thus, $b_1, b_2$ are constrained to be 
\ie\label{bonetwo}
b_1(x) = - { a(x_1)\over x-x_1} ,~~~~
b_2(x) = - { a(x_2) \over x-x_2}.
\fe
The non-vanishing of $b_2(x)$ at $x_1$, however, introduces a singularity that must be cancelled by a pole of $c(x)$ at $x_1$.
Indeed, it follows from (\ref{wetachfir}) that
\ie
\nu_2 [\dif w_1 | \dif \eta_1] 
&\sim \frac{f'(x)}{2p(x)}\left(\alpha^{-1} \nu_2 +{f'(x)\over 2y}\frac{\eta \nu_1 \nu_2}{x - x_1}\right)[\dif x| \dif \eta]
\fe
near $x_1$. Comparing with (\ref{jexpr}), we find the residue of $c(x)$ at $x_1$, $c^{(-1)}(x_1) = {f'(x_1)\over 2} b_2(x_1)$. A similar analysis near $x_2$ gives $c^{(-1)}(x_2) = - {f'(x_2)\over 2} b_1(x_2)$. Combining with (\ref{bonetwo}), and the behavior at infinity, we conclude that
\ie
c(x) = - {f'(x_1) a(x_2)\over 2 x_{12} (x-x_1)} - {f'(x_2) a(x_1)\over 2 x_{12} (x-x_2)} +\wt c(x), 
\fe
where $\wt c(x)$ is a linear polynomial.

Let us turn to the split SRS $\mathfrak{C}'$, parameterized by the supercoordinates $(x',\eta')$, whose underlying hyperelliptic curve is 
\ie
y'^2 = \wt f(x')=\wt p(x') \wt q(x'),
\fe 
with fermionic derivative 
\ie
D_{\eta'} = \partial_{\eta'} + {\wt f'(x')\over 2\wt p(x)} \eta' \partial_{y'}.
\fe
The fermionic coordinate $\tau'$ is related by the transition map $\tau' = (\wt p(x'))^{-1} y'\eta'$.
We will assume that $\wt f$ is related to $f$ by the infinitesimal shift
\ie
\wt f(x') = f(x') - m(x') \nu_1 \nu_2 = (p(x') + n(x')\nu_1 \nu_2)(q(x') + t(x')\nu_1 \nu_2).
\fe 
Now suppose that there is an integral form $J'$ on $\mathfrak{C}'$,
\ie\label{jprimedef}
J' = \eta' y'^{-1} \wt a(x') [\dif x'|\dif  \eta'],
\fe
where $\wt a(x')$ is linear in $x'$, that has the same periods as those of $J$ on $\mathfrak{C}_\nu$. We would like to determine $m(x')$, and the relation between $\wt a(x')$ and $a(x)$.

To calculate the periods of $J'$, it will be convenient to define unprimed coordinates $x,y,\eta,\tau$, related by
\ie\label{eqn:srs period coordinate matching}
&x' = x, ~~~~~ y' = y - \frac{m(x)}{2 y} \nu_1 \nu_2, \\
&\eta' = \left[ 1 + \frac{1}{4} \left(\frac{n(x)}{p(x)} - \frac{t(x)}{q(x)}\right) \nu_1\nu_2 \right] \eta, ~~~~~ \tau' = \left[1 - \frac{1}{4} \left(\frac{n(x)}{p(x)} - \frac{t(x)}{q(x)}\right) \nu_1\nu_2 \right] \tau,
\fe
that obey $y^2=f(x)$ and $\tau=(p(x))^{-1} y \eta$. The fermionic derivative in the unprimed coordinates is $D_\eta = \partial_\eta + {f'(x)\over 2p(x)} \eta \partial_y$. Note that the coordinate transformation (\ref{eqn:srs period coordinate matching}) is (necessarily) singular at the branch points and at infinity, but this will not affect the calculation of periods which can be defined by integration along paths that avoid the singular points. In the unprimed coordinates, (\ref{jprimedef}) is written as
\ie\label{jprunpr}
J' = \eta y^{-1} \left(1 + \frac{m(x)}{2 f(x)}\nu_1 \nu_2\right) \wt a(x) [\dif x| \dif \eta].
\fe
The requirement that $J$ (\ref{jexpr}) has the same periods as $J'$ amounts to
\ie\label{eqn:srs total deriv constraint}
& y^{-1} \left[ a(x) + c(x)\nu_1\nu_2 - \left(1 + \frac{m(x)}{2 f(x)}\nu_1 \nu_2\right) \wt a(x)  \right] \dif x = \dif \left( {u(x) + v(x) \nu_1\nu_2 \over y}\right),
\fe
for some polynomials $u(x)$ and $v(x)$. It follows that
\ie\label{afcond}
& \wt a(x) = a(x) + \ell(x) \nu_1\nu_2,
\\
& f(x) \left( c(x) - \ell(x) \right) - {1\over 2} m(x) a(x) = f(x) v'(x) - {1\over 2} f'(x) v(x),
\fe
where $\ell(x)$ is linear in $x$, and $u(x)$ necessarily vanishes. 
At the branch points $x_i$ for $i\not=1,2$, the condition reduces to
\ie
m(x_i) a(x_i) = f'(x_i) v(x_i)
\fe
At $x_1$ and $x_2$, we have
\ie
m(x_1) a(x_1) &= -\frac{\left(f'(x_1)\right)^2}{x_{12}} a(x_2) + f'(x_1) v(x_1), 
\\
m(x_2) a(x_2) &= -\frac{\left(f'(x_2)\right)^2}{x_{12}} a(x_1) + f'(x_2) v(x_2).
\fe 
Let us define
\ie\label{mvzero}
v_0(x) = - {f(x)\left(a(x_1) - a(x_2)\right) \over (x - x_1) (x - x_2)} ,~~~~ m_0(x) = - {f(x)\over x_{12}} \left(\frac{f'(x_1)}{x - x_1} + \frac{f'(x_2)}{x - x_2}\right) ,
\fe
and
\ie
v(x) \equiv v_0(x) + r(x),~~~~ m(x) = m_0(x) + h(x).
\fe
The condition (\ref{afcond}) at all branch points can be expressed as
\ie\label{eqn:srs period branchpoints}
h(x_i) a(x_i) = f'(x_i) r(x_i).
\fe
The LHS of (\ref{afcond}) has degree at most 7, and it follows that $v(x)$ has degree at most 3.\footnote{Note that in this case, the degree 8 terms on the RHS of (\ref{afcond}) cancel.} Since we have separated from $v(x)$ the degree 4 polynomial $v_0(x)$, $r(x)$ must also have degree 4. In particular, $r(x)$ is fixed by its value at the branch points, determined through (\ref{eqn:srs period branchpoints}).

We can use the Euclidean algorithm to decompose 
\ie\label{hresults}
h(x) = A(x) f'(x) + B(x) f(x),
\fe
where $A(x)$ has degree at most 5, and write (\ref{eqn:srs period branchpoints}) equivalently as
\ie\label{axrcond}
A(x_i) a(x_i) = r(x_i).
\fe
Since $A(x)a(x)$ has degree at most 6, it is fixed by (\ref{axrcond}) to be
\ie
A(x) a(x) = r(x).
\fe
It follows that $A(x)$ in fact has degree 3, and takes the form
\ie\label{aasth}
A(x) = A_0 + A_1 x + A_2 x^2 + x_{12} f^{(6)} x^3.
\fe 
The branch points of $\wt f(x)$ are
\ie
x_i' = x_i - s_i \nu_1\nu_2,~~~~{\rm with}~~
s_i = - {m(x_i)\over f'(x_i)}.
\fe
The result for $m_0$ and $h(x)$ given by (\ref{mvzero}) and (\ref{hresults}), (\ref{aasth}) then gives (\ref{sresulta}).

\bibliographystyle{JHEP}
\bibliography{srs-g2}

\providecommand{\href}[2]{#2}\begingroup\raggedright\begin{thebibliography}{10}

\bibitem{DHoker:1988pdl}
E.~D'Hoker and D.~H. Phong, {\it {The Geometry of String Perturbation Theory}},
   {\em Rev. Mod. Phys.} {\bf 60} (1988) 917.

\bibitem{Witten:2012ga}
E.~Witten, {\it {Notes On Super Riemann Surfaces And Their Moduli}},  {\em Pure
  Appl. Math. Quart.} {\bf 15} (2019), no.~1 57--211,
  [\href{http://arxiv.org/abs/1209.2459}{{\tt arXiv:1209.2459}}].

\bibitem{Friedan:1985ge}
D.~Friedan, E.~J. Martinec, and S.~H. Shenker, {\it {Conformal Invariance,
  Supersymmetry and String Theory}},  {\em Nucl. Phys. B} {\bf 271} (1986)
  93--165.

\bibitem{Verlinde:1987sd}
E.~P. Verlinde and H.~L. Verlinde, {\it {Multiloop Calculations in Covariant
  Superstring Theory}},  {\em Phys. Lett. B} {\bf 192} (1987) 95--102.

\bibitem{Sen:2014pia}
A.~Sen, {\it {Off-shell Amplitudes in Superstring Theory}},  {\em Fortsch.
  Phys.} {\bf 63} (2015) 149--188, [\href{http://arxiv.org/abs/1408.0571}{{\tt
  arXiv:1408.0571}}].

\bibitem{Sen:2015hia}
A.~Sen and E.~Witten, {\it {Filling the gaps with PCO\textquoteright{}s}},
  {\em JHEP} {\bf 09} (2015) 004, [\href{http://arxiv.org/abs/1504.00609}{{\tt
  arXiv:1504.00609}}].

\bibitem{Wang:2022zad}
C.~Wang and X.~Yin, {\it {On the Equivalence between SRS and PCO Formulations
  of Superstring Perturbation Theory}},
  \href{http://arxiv.org/abs/2205.01106}{{\tt arXiv:2205.01106}}.

\bibitem{DHoker:2001kkt}
E.~D'Hoker and D.~H. Phong, {\it {Two loop superstrings. 1. Main formulas}},
  {\em Phys. Lett. B} {\bf 529} (2002) 241--255,
  [\href{http://arxiv.org/abs/hep-th/0110247}{{\tt hep-th/0110247}}].

\bibitem{DHoker:2001qqx}
E.~D'Hoker and D.~H. Phong, {\it {Two loop superstrings. 2. The Chiral measure
  on moduli space}},  {\em Nucl. Phys. B} {\bf 636} (2002) 3--60,
  [\href{http://arxiv.org/abs/hep-th/0110283}{{\tt hep-th/0110283}}].

\bibitem{DHoker:2001foj}
E.~D'Hoker and D.~H. Phong, {\it {Two loop superstrings. 3. Slice independence
  and absence of ambiguities}},  {\em Nucl. Phys. B} {\bf 636} (2002) 61--79,
  [\href{http://arxiv.org/abs/hep-th/0111016}{{\tt hep-th/0111016}}].

\bibitem{DHoker:2001jaf}
E.~D'Hoker and D.~H. Phong, {\it {Two loop superstrings 4: The Cosmological
  constant and modular forms}},  {\em Nucl. Phys. B} {\bf 639} (2002) 129--181,
  [\href{http://arxiv.org/abs/hep-th/0111040}{{\tt hep-th/0111040}}].

\bibitem{DHoker:2002hof}
E.~D'Hoker and D.~H. Phong, {\it {Lectures on two loop superstrings}},  {\em
  Conf. Proc. C} {\bf 0208124} (2002) 85--123,
  [\href{http://arxiv.org/abs/hep-th/0211111}{{\tt hep-th/0211111}}].

\bibitem{DHoker:2005dys}
E.~D'Hoker and D.~H. Phong, {\it {Two-loop superstrings. V. Gauge slice
  independence of the N-point function}},  {\em Nucl. Phys. B} {\bf 715} (2005)
  91--119, [\href{http://arxiv.org/abs/hep-th/0501196}{{\tt hep-th/0501196}}].

\bibitem{DHoker:2005vch}
E.~D'Hoker and D.~H. Phong, {\it {Two-loop superstrings VI: Non-renormalization
  theorems and the 4-point function}},  {\em Nucl. Phys. B} {\bf 715} (2005)
  3--90, [\href{http://arxiv.org/abs/hep-th/0501197}{{\tt hep-th/0501197}}].

\bibitem{DHoker:2007csw}
E.~D'Hoker and D.~H. Phong, {\it {Two-Loop Superstrings. VII. Cohomology of
  Chiral Amplitudes}},  {\em Nucl. Phys. B} {\bf 804} (2008) 421--506,
  [\href{http://arxiv.org/abs/0711.4314}{{\tt arXiv:0711.4314}}].

\bibitem{DHoker:2020prr}
E.~D'Hoker, C.~R. Mafra, B.~Pioline, and O.~Schlotterer, {\it {Two-loop
  superstring five-point amplitudes. Part I. Construction via chiral splitting
  and pure spinors}},  {\em JHEP} {\bf 08} (2020) 135,
  [\href{http://arxiv.org/abs/2006.05270}{{\tt arXiv:2006.05270}}].

\bibitem{DHoker:2020tcq}
E.~D'Hoker, C.~R. Mafra, B.~Pioline, and O.~Schlotterer, {\it {Two-loop
  superstring five-point amplitudes. Part II. Low energy expansion and
  S-duality}},  {\em JHEP} {\bf 02} (2021) 139,
  [\href{http://arxiv.org/abs/2008.08687}{{\tt arXiv:2008.08687}}].

\bibitem{DHoker:2021kks}
E.~D'Hoker and O.~Schlotterer, {\it {Two-loop superstring five-point
  amplitudes. Part III. Construction via the RNS formulation: even spin
  structures}},  {\em JHEP} {\bf 12} (2021) 063,
  [\href{http://arxiv.org/abs/2108.01104}{{\tt arXiv:2108.01104}}].

\bibitem{Takayanagi:2003sm}
T.~Takayanagi and N.~Toumbas, {\it {A Matrix model dual of type 0B string
  theory in two-dimensions}},  {\em JHEP} {\bf 07} (2003) 064,
  [\href{http://arxiv.org/abs/hep-th/0307083}{{\tt hep-th/0307083}}].

\bibitem{Douglas:2003up}
M.~R. Douglas, I.~R. Klebanov, D.~Kutasov, J.~M. Maldacena, E.~J. Martinec, and
  N.~Seiberg, {\it {A New hat for the c=1 matrix model}},  in {\em {From Fields
  to Strings: Circumnavigating Theoretical Physics: A Conference in Tribute to
  Ian Kogan}}, pp.~1758--1827, 7, 2003.
\newblock \href{http://arxiv.org/abs/hep-th/0307195}{{\tt hep-th/0307195}}.

\bibitem{Balthazar:2022atu}
B.~Balthazar, V.~A. Rodriguez, and X.~Yin, {\it {The S-Matrix of 2D Type 0B
  String Theory Part 1: Perturbation Theory Revisited}},
  \href{http://arxiv.org/abs/2201.05621}{{\tt arXiv:2201.05621}}.

\bibitem{Balthazar:2022apu}
B.~Balthazar, V.~A. Rodriguez, and X.~Yin, {\it {The S-Matrix of 2D Type 0B
  String Theory Part 2: D-Instanton Effects}},
  \href{http://arxiv.org/abs/2204.01747}{{\tt arXiv:2204.01747}}.

\end{thebibliography}\endgroup

\end{document}